\documentclass[11pt,a4paper]{article}

\newcommand{\bbox}[1]{\boldsymbol{#1}}
\newcommand{\CC}{{\cal C}}
\newcommand{\LL}{{\cal L}}
\newcommand{\FF}{{\cal F}}
\newcommand{\DD}{{\cal D}}
\newcommand{\Tr}{\mathop{\mathrm{Tr}}}

\usepackage{amsmath}
\usepackage{graphics}
\usepackage{psfrag}
\usepackage{array}		
\usepackage{legendes}
\numberwithin{equation}{section}

\def \dd{\hbox{d}}

\begin{document}
\title{Screened cluster expansions for partially ionized gases}

\author{A. Alastuey\thanks{Laboratoire de physique, \'Ecole Normale Sup\'erieure de Lyon, UMR 5672 du CNRS, 46, all\'ee d'Italie, 69364 Lyon Cedex 07, \textsc{France}.}\ ,
V. Ballenegger\thanks{
Institut de Th\'eorie des Ph\'enom\`enes Physiques
\'Ecole Polytechnique F\'ed\'erale de Lausanne EPFL,
CH-1015 Lausanne, \textsc{Switzerland}.}\ \thanks{Present address: Department of Chemistry, University of Cambridge, Lensfield Road, Cambridge CB2 1EW, \textsc{United Kingdom}, E-mail: vcb25@cam.ac.uk.}\ , F. Cornu\thanks{Laboratoire de Physique Th\'eorique, Universit\'e Paris-Sud,
UMR 8627 du CNRS, B\^{a}timent 210, 91405 Orsay, \textsc{France}.}\ \ and Ph. A. Martin$^\dag$
}

\date{\today}

\maketitle     

\begin{abstract}
We consider a partially ionized gas at thermal equilibrium, in the Saha regime. The system is described in terms of a quantum plasma of nuclei and electrons. In this framework, the Coulomb interaction is the source of a large variety of phenomena occuring at different scales: recombination, screening, diffraction, etc. In this paper, we derive a cluster expansion adequate for a coherent treatment of those phenomena. The expansion is obtained by combining the path integral representation of the quantum gas with familiar Mayer diagrammatics. In this formalism, graphs have a clear physical interpretation: vertices are associated with recombined chemical species, while bonds describe their mutual interactions. The diagrammatical rules account exactly for all effects in the medium. Applications to thermodynamics, van der Waals forces and dielectric versus conductive behaviour will be presented in forthcoming papers.
\end{abstract}


\section{Introduction}

This paper is devoted to the construction of a cluster expansion adapted to the study of partially ionized gases. In the so-called physical picture of such gases, one considers an assembly of non relativistic point nuclei and electrons in thermal equilibrium with appropriate statistics and interacting with the Coulomb potential. That fundamental approach is to be contrasted with the chemical picture which introduces preformed chemical species as elementary constituents. At this basic level of description all characteristic phenomena of charged systems such as atomic or molecular recombination, ionization, ionic and dielectric screening originate from the sole Coulomb interaction between charges. It is notoriously difficult to treat all of them from first principles in a fully coherent way. In particular one has to deal simultaneously and without intermediate modelization with the short distance part of 
Coulomb potential which gives rise to chemical binding 
(at the scale of Bohr radius), and with its long range part responsible 
for collective screening effects (at the scale of the classical Debye length).

An important point should be stressed straight away. The very concept 
of atom and molecule can only make sense in a low density or moderately 
dense medium and at temperatures sufficiently low to be away 
from full ionization. This situation is captured in a scaling limit 
called the atomic (or molecular) limit, which is more precisely described in Section 5 in the case of the Hydrogen plasma. In that limit, one lets temperature and density tend to zero in a coupled way. Low temperature 
favors binding over ionization, whereas low density, by increasing the 
available phase space, favors dissociation. The rate at which 
density is reduced as the temperature tends to zero determines 
an energy entropy balance that selects the formation of some 
chemical species. In the strict limit one recovers 
the thermodynamical laws of mixtures of ideal substances.  
Thus the proper physics of partially ionized gases lies in the 
vicinity of the atomic limit: this defines the Saha regime for which 
it is of interest to provide a suitable expansion.

In the literature, there exists various attempts, within the physical picture, 
to derive suitable expansions at low temperature and low density. 
First, the effective-potential method \cite{Mor} has been applied to 
quantum plasmas by Ebeling \cite{Ebe}. That first-principles approach 
amounts to introduce an equivalent classical system of point particles with 
many-body potentials which incorporate all quantum effects.
Low density expansions at fixed non-zero temperature have been 
successfully computed \cite{Ebe,KraKreEbeRop,Kah}, while 
plausible expressions for basic ingredients of the chemical picture (like 
effective fugacities of recombined entities) have also been proposed 
within that framework \cite{KraKreEbeRop,EbeKraKre}. Unfortunately, an 
explicit control of the contribution arising from any $n$-body potential
($n \geq 3$) cannot be achieved by using simple diagrammatical techniques 
(in most previous works, only two-body potentials are retained). Therefore, 
all effects related to recombination into complex entities 
of more than two particles, cannot be properly taken into account. 
Another approach \cite{Rog1} relies on a heuristic 
introduction of quantum and 
screening effects in classical versions of usual fugacities 
expansions. Roughly speaking, Boltzmann factors at short distances are 
replaced by their quantum counterparts, while Coulomb interactions at large distances are exponentially screened over classical Debye screening length. 
If that method may provide rather good quantitative results at moderate temperatures and densities \cite{Rog2}, explicit expressions 
for contributions related to any phenomenon are not available. 
Eventually, we emphasize that many-body 
perturbation theory \cite{FetWal,MarRot} is not suited to the 
description of low temperature and low density regimes. 
Indeed, perturbative expansions with respect to the coupling constant 
(the electric charge) cannot account for recombination where 
interactions play a crucial non-perturbative role. Consequently, that 
formalism is more appropriate for dealing with fully ionized regimes, either 
at low temperature and high density \cite{MonWar}, or 
at high temperature and low density \cite{Dew,DewSchSakKra}. 

In order to circumvent the drawbacks of previous methods, we introduce 
the path integral representation of quantum Coulomb gases. As shown 
in this paper, that formalism allows us to build an 
expansion which fulfills all requirements for a proper description of 
Saha regime. In particular, collective screening effects, as well as 
quantum mechanics of arbitrary particle clusters, are simultaneously 
dealt with. For the sake of presentation simplicity, derivations are carried out 
for the Hydrogen plasma made with protons and electrons (see Section 2). 
Extensions to other quantum gases are briefly mentionned in Section 5.

The functional integral representation of quantum Coulomb gas by means of Feynman-Kac formula \cite{SimSchRoe} is displayed in Section 3.1. Charges of the same kind and same statistics are collected according to the cycle decomposition 
of the permutation group into closed Brownian paths called charged loops \cite{Gin}. In terms of those loops, statistical averages are performed according to the same rules as in classical statistical mechanics. 
As a consequence, the powerful methods of Mayer graphs are available 
in the space of loops \cite{Cor1}: activities including the self-energy of a loop 
are associated to vertices, and bonds are usual Mayer factors of 
interaction between different loops. Because of the long range of the Coulomb potential, those bonds are not integrable, and one has first to 
resum classes of diverging graphs in order to obtain an effective potential 
with shorter range.  Since details have already been worked out in previous works~\cite{AlaCorPer,Cor1} we merely summarize the procedure in section 3.2 for the sake of completeness. The procedure follows 
the familiar chain graph summation as in the classical case 
\cite{MaySal,Mee,Abe}. The subsequent effective 
potential is the quantum analogue of the Debye potential. It incorporates specific features of quantum mechanics, namely quantum statistics, short distance Gibbs weight leading to binding, 
as well as algebraically decaying tails due to quantum fluctuations. 
That effective potential is extensively studied in~\cite{BalMarAla}: 
with its help one can define a resummed diagrammatics of 
prototype graphs that are free from Coulomb divergencies. 
The rules are the same as those of standard Abe-Meeron diagrammatics. 
This fugacity expansion in terms of loop prototype graphs is suited 
to derive, at fixed non-zero temperature, low density virial 
equation of state in the ionized phase~\cite{AlaPer,BryMar}. 
The small parameter is the density $\rho_{e}$ of unbound electrons 
so that contributions of bound states (e.g. hydrogen atoms) only occur 
as small corrections at order $\rho^{2}_{e}$.

The expansion (\ref{quatrequinze}) in terms of loop prototype graphs 
is not well adapted to a direct analysis of Saha regime, 
where recombined atoms or molecules appear with comparable densities 
at low temperature. The reason for this inadequacy is that 
Gibbs factors arising in a given prototype graph do not provide 
the complete set of pair interactions that would correspond 
to an atomic or molecular Hamiltonian. Hence each prototype 
graph is not in direct relation with the quantities 
that are naturally involved in the definitions of energies and 
densities of recombined entities. Then, the aim is to systematically 
reorganize the previous loop expansion into an expansion where clusters of charges are fully interacting 
and are in one to one correspondance 
with the possible chemical binding processes. This is the content of 
Section 4, where we derive the main result of the paper, namely new form (\ref{cinqvingt}) of the fugacity expansion that we call the screened cluster expansion.

The derivation is accomplished in three steps. The first one (Section 4.1)
consists in collecting prototype graphs into classes such that
the sum of graphs in a class builds weights of fully interacting 
clusters of loops. This provides Gibbs factors
where all pairwise interactions between loops are present. 
The building rules of those coefficients are the same as 
those of usual Mayer cofficients in standard fugacity expansions. 
In fact, had one to deal with a gas of particles interacting by 
means of short range forces, those coefficients would simply be 
identical to standard Mayer coefficients written within 
Feynman-Kac functional integral formalism. The difference here is 
that loops interact with the effective potential requested by 
the long range of Coulomb interaction. As that effective potential itself 
results from chain summations, one must be careful to avoid double counting 
of graphs. It turns out that avoiding double counting induces 
residual interactions between clusters through the effective potential. 
If now Mayer coefficients of fully interacting clusters are 
interpreted as dressed activities associated with vertices 
and residual interactions are associated with bonds, it is nice 
that the resulting diagrammatics is still of Mayer type in terms of 
those weights and bonds: this leads to formally exact representation \eqref{cinqquatre} of the loop density. At any non-zero temperature,
each graph of (\ref{cinqquatre}) is finite since bonds and vertices 
are entirely constructed with the help of the effective potential. 

The necessity of the second step performed in Section 4.2 is the following.
As explained before, the physical range of thermodynamical parameters 
relevant for the description of partially ionized gases lies in 
the neighborhood of the atomic limit, namely for vanishing densities and temperatures. If one lets the density tend to zero in weights of 
(\ref{cinqquatre}), the effective potential reduces to the bare Coulomb 
potential so that one faces again diverging spatial integrals. 
Hence one needs to introduce a further step if one wants 
cluster integrals to have a well defined atomic limit. In rough terms, 
it consists simply in a regularization procedure of divergent 
integrals by substraction of non-integrable part of their integrands. 
This leads to the introduction of truncated weights in such a way that 
new weights remain integrable when screening is removed. 
The definition of truncated cluster weights adopted here has a 
remarkable property: it preserves the topological rules of 
Mayer diagrammatics. This leads to expansion (\ref{cinqseize}) 
for the loop density. The vertices are truncated Mayer coefficients, 
and a bond between two clusters links every loop in one cluster to every loop in the other cluster \textsl{via} the effective potential (or its second and third power). 

In a third step (Section 4.3), we introduce particle clusters. If in each graph of (\ref{cinqseize}), we detail phase-space 
integrations over loop internal degrees of freedom, loop clusters give rise 
to particle clusters. The final step amounts to collect all contributions 
with the same topological structure in terms of particle clusters. Then, we obtain the so-called screened cluster 
expansion (\ref{cinqvingt}) for the particle density. 
In that diagrammatic series, the graphs $G$ (see Fig.12) are identical to familiar 
Mayer graphs with particle clusters in place of one-particle points.   
The internal state of a cluster is determined by a 
distribution of its particles into a given set of loops. Its statistical weight 
reduces to the truncated Mayer coefficient for that loop configuration. 
Two clusters are connected by at most one bond, which can be either 
the total screened interaction between the respective two sets of 
particles, or the square or the cube of that potential. The topological 
rules are usual Mayer rules, except for some specific rules that avoid 
double counting (see Section 4.4). Screened cluster expansions 
for other equilibrium quantities are discussed in Section 4.5. In particular, 
formula (\ref{cinqvingthuit}) allows one to study particle correlations.

As detailed in Section 5, screened cluster expansions do incorporate physics required for treating 
Saha regime. All Coulomb divergencies are removed \textsl{via} the 
introduction of screened interactions with proper quantum fluctuations. 
Recombination into complex entities is accounted for by particle clusters, 
because truncated Mayer coefficients do preserve the non-perturbative 
structure of genuine Gibbs operators with respect to interactions on one hand, while summation over all associated loop configurations restore 
correct Fermi symmetrization on another hand. Moreover, since the screened 
interaction is close to the bare Coulomb potential at short distances, 
familiar chemical species defined in the vacuum do emerge, for 
instance atoms $H$, molecules $H_2$, ions $H^-$ and $H_2^+$, etc...   
Statistical weights of particle clusters are in correspondance with the densities of the associated chemical species formed by 
recombination of $N_{p}$ protons and $N_{e}$ electrons. 
Such weights incorporate thermal excitations, thermal 
and pressure dissociation, as well as 
spectral broadening of energy levels by collective effects.
Bonds between clusters describe all possible interactions between substances present in the gas.

The present paper is merely devoted to the presentation of the general formalism. In Section 5, we also discuss a few possible applications. In forthcoming papers, we shall treat quantitatively several physical situations of interest, in particular non ideal contributions to Saha equation of state \cite{AlaBalCorMar}, the dielectric response of an atomic hydrogen gas \cite{BalMar} and van der Waals forces in a partially recombined plasma \cite{AlaCorMar}.

\section{The Hydrogen plasma}

\subsection{Definition of the model}

We consider the two-component system (in three dimensions) made of protons and electrons. In the present non-relativistic limit, the proton and the electron are viewed as quantum point particles with respective masses, charges and spins, $e_p=e$ and $e_e=-e$, $m_p$ and $m_e$, $\sigma_p=1/2$ and $\sigma_e=1/2$. The kinetic energy operator for each particle of species $\alpha=p,e$ with position $\textbf{x}$ takes the Schrodinger form $-{\hbar^2 \over 2m_{\alpha}} \Delta$ where $\Delta$ is the Laplacian with respect to $\textbf{x}$. Two particles separated by a distance $r$ interact via the instantaneous Coulomb potential $v(r)=1/r$. The corresponding Coulomb Hamiltonian $H_{N_p,N_e}$ for $N_p$ protons and $N_e$ electrons reads
\begin{equation}
H_{N_p,N_e} = -\sum_{i=1}^N {\hbar^2 \over 2m_{\alpha_i}} \Delta_i + {1 \over 2}\sum_{i \neq j} e_{\alpha_i} e_{\alpha_j} v(|\textbf{x}_i - \textbf{x}_j|)
\label{deuxun}
\end{equation}
where $N=N_p+N_e$ is the total number of particles. In (\ref{deuxun}), the subscript $i$ is attached to protons for $i=1,...,N_p$ and to electrons for $i=N_p+1,...,N_p+N_e$, so the species index $\alpha_i$ reduces either to $p$ or $e$ while $\textbf{x}_i$ denotes either the position $\textbf{R}_i$ of the $i$-th proton or the position $\textbf{r}_j$ of the $j$-th electron ($j=i-N_p$).

The system is enclosed in a box with volume $\Lambda$, in contact with a thermostat at temperature $T$ and a reservoir of particles that fixes the chemical potentials equal to $\mu_p$ and $\mu_e$ for protons and electrons respectively. Its grand-partition function $\Xi$ is ($\beta=1/(k_B T)$)
\begin{equation}
\Xi= \Tr \exp \left[-\beta (H_{N_p,N_e} - \mu_p N_p - \mu_e N_e)\right].
\label{deuxdeux}
\end{equation}
In (\ref{deuxdeux}), the trace is taken over all states symmetrized according to the Fermionic nature of each species; the boundary conditions for the wave functions at the surface of the box can be chosen of the Dirichlet type. Lieb and Lebowitz \cite{LieLeb} have proved that the thermodynamic limit ($\Lambda \rightarrow \infty$ at fixed $\beta$ and $\mu_{\alpha}$) exists, thanks to Fermi statistics and screening. Indeed, the Fermionic statistics of at least one species implies the $H$-stability \cite{DysLen} 
\begin{equation}
H_{N_p,N_e} > -B (N_p + N_e),  \qquad B>0
\label{deuxtrois}
\end{equation}
that prevents the collapse of the system. On the other hand, screening ensures that it does not explode. In a fluid phase, the infinite system maintains local neutrality, i.e. the homogeneous local particle densities $\rho_p$ and $\rho_e$ for protons and electrons remain equal for any choice of the chemical potentials $\mu_{\alpha}$. In other words, the common particle density $\rho=\rho_p=\rho_e$, as well as all other bulk equilibrium quantities, depend on the sole combination
\begin{equation}
\mu=(\mu_p + \mu_e)/2, 
\label{deuxquatre}
\end{equation}
and not on the difference $\nu=(\mu_e-\mu_p)/2$. In particular, in terms of the fugacities $z_{\alpha}= \exp (\beta \mu_{\alpha})$, this means that both the density $\rho$ and the pressure $P$ are functions of $\beta$ and $z=(z_p z_e)^{1/2}= \exp (\beta \mu)$ only. In the following, we set 
\begin{equation}
\mu_p=\mu - {3 \over 2}k_B T \ln {\lambda_e \over \lambda_p},\qquad  \mu_e=\mu + {3 \over 2}k_B T \ln{\lambda_e \over \lambda_p}
\label{deuxcinq}
\end{equation} 
where $\lambda_{\alpha}=( \beta \hbar^2 / m_{\alpha})^{1/2}$ is the thermal de Broglie wavelength of species $\alpha$. This choice guarantees that the Maxwell-Boltzmann densities $\rho_p^{id} = 2z_p / (2\pi \lambda_p^2)^{3/2}$ and $\rho_e^{id} = 2z_e / (2\pi \lambda_e^2)^{3/2}$ of free (no interactions) proton and electron gases respectively, are identical, i.e. $\rho_p^{id}= \rho_e^{id}= 2z/  (2\pi \lambda^2)^{3/2}$ with
$\lambda = (\lambda_p  \lambda_e)^{1/2}$. The enforced neutrality
\begin{equation}
\sum_{\alpha} e_{\alpha}  z_{\alpha} / (2\pi \lambda_{\alpha}^2)^{3/2} = 0
\label{deuxsix}
\end{equation}
of the ideal mixture simplifies the analysis of the screened cluster 
expansions for the interacting system derived in Section 4.

\subsection{Formal Mayer expansions}

For describing dilute regimes at low fugacities ($z \ll 1$), 
Mayer fugacity expansions are \textsl{a priori} well appropriate. That 
expansion for the pressure is easily inferred, at a formal level, from 
identity 
\begin{equation}
\beta P = {\ln \Xi \over \Lambda},
\label{troisun}
\end{equation}
where it is understood that thermodynamic limit 
$\Lambda \rightarrow \infty$ is taken once for all. They read
\begin{equation}
\beta P = \sum_{(N_p,N_e) \neq (0,0)} {z_p^{N_p}z_e^{N_e} \over N_p!N_e!} B_{N_p,N_e} 
\label{troisdeux}
\end{equation}
where Mayer coefficients $B_{N_p,N_e}$ can be expressed as suitable traces, 
\begin{equation}
B_{N_p,N_e} = {1 \over \Lambda}\Tr \left[\exp (-\beta H_{N_p,N_e})\right]_{\text{Mayer}}.
\label{troisdeuxa}
\end{equation}
The first Mayer operators $[\exp (-\beta H_{N_p,N_e})]_{\text{Mayer}}$ read
\begin{gather}	\notag
[\exp (-\beta H_{1,0})]_{\text{Mayer}} = \exp (-\beta H_{1,0}), \qquad
[\exp (-\beta H_{0,1})]_{\text{Mayer}} = \exp (-\beta H_{0,1}), \\
[\exp (-\beta H_{1,1})]_{\text{Mayer}} = \exp (-\beta H_{1,1}) - \exp (-\beta H_{1,0})  \exp (-\beta H_{0,1}),
\label{troistrois}
\end{gather}
while similar expressions can be obtained for 
$[\exp (-\beta H_{N_p,N_e})]_{\text{Mayer}}$ by considering all possible partitions 
of $N_p$ protons and $N_e$ electrons. The traces (\ref{troisdeuxa}) 
must be taken over symmetrized states according to Fermi statistics: for each 
product of Gibbs operators $\exp (-\beta H_{M_p,M_e})$ ($M_p \leq N_p$, 
$M_e \leq N_e$) in $[\exp (-\beta H_{N_p,N_e})]_{\text{Mayer}}$, such states are
products of symmetrized states made with $M_p$ protons and $M_e$ electrons. 
For instance, in space of positions and spins, $B_{2,0}$ reads
\begin{multline}
B_{2,0} = {1 \over \Lambda} \int \dd \textbf{R}_1 \int \dd \textbf{R}_2 
\big[4\langle \textbf{R}_1 \textbf{R}_2|\exp (-\beta H_{2,0})
|\textbf{R}_1 \textbf{R}_2\rangle \\ 
- 4\langle \textbf{R}_1|\exp (-\beta H_{1,0})|\textbf{R}_1\rangle
\langle \textbf{R}_2|\exp (-\beta H_{1,0})|\textbf{R}_2\rangle \\
-2\langle \textbf{R}_2 \textbf{R}_1|
\exp (-\beta H_{2, 0})|\textbf{R}_1 \textbf{R}_2\rangle\big]
\label{troistroisa}
\end{multline}
(the prefactors 4 and 2 are due to spin degeneracy).

Mayer series for particle densities $\rho_{p,e}$, are easily obtained 
by inserting (\ref{troisdeux}) 
into the identities
\begin{equation}
\rho_{\alpha} = z_{\alpha} {\partial \beta P \over \partial z_{\alpha}}
\label{troisquatre}
\end{equation}
for $\alpha = p,e$. These series read
\begin{equation}
\rho_p = \sum_{N_p=1,N_e=0}^{\infty} {z_p^{N_p}z_e^{N_e} \over (N_p-1)!N_e!} B_{N_p,N_e} 
\label{troiscinq}
\end{equation}
and
\begin{equation}
\rho_e = \sum_{N_p=0,N_e=1}^{\infty} {z_p^{N_p}z_e^{N_e} \over N_p!(N_e-1)!} B_{N_p,N_e} 
\label{troissix}.
\end{equation} 
Similar Mayer series can be derived for any equilibrium quantity. 
Contrary to the case of systems with short range forces, all 
these series must be understood in a formal sense:
Mayer coefficients $B_{N_p,N_e}$ diverge because the Coulomb potential is not integrable at large distances.
The first step of our method consists in removing all  
Coulomb divergencies \textsl{via} systematic resummations of chain 
interactions in quantum Mayer graphs defined for an equivalent gas of 
classical loops (see Section 3). Then, a suitable reorganization 
of the resulting resummed graphs provides the required 
screened cluster expansion which is well suited for describing 
both low temperature and low fugacity regimes (see Section 4). In particular, 
Mayer series (\ref{troiscinq}) are rewritten in terms of well-behaved 
resummed coefficients $B_{N_p,N_e}^{(R)}$, that do account for 
recombination into complex entities at low temperature (see Section 5).   

\section{Path integral formalism}

Within the Feynman-Kac path integral representation, we introduce the equivalent classical gas of charged loops (section 3.1). Mayer fugacity expansions for this sytem are considered. We briefly sketch the resummation scheme which removes all long-range Coulomb divergencies in the Mayer graphs. This provides
resummed diagrammatic expansions of the equilibrium quantities in terms of 
a well-behaved screened potential.

\subsection{The gas of charged loops}

In the definition (\ref{deuxdeux}) of the grand-partition function, the trace can be taken over symmetrized Slater sums built with one-body states $|\textbf{x}\sigma^z\rangle$ that describe a particle localized at $\textbf{x}$ with the $z$-component of its spin equal to $\sigma^z$. This provides

\begin{multline}
\Xi = \sum_{N_p, N_e=0}^{\infty} { z_p^{N_p}z_e^{N_e} \over N_p!N_e!} \sum_{{\cal P}_p,{\cal P}_e} \epsilon ({\cal P}_p)\epsilon ({\cal P}_e) \sum_{\{\sigma_{p,i}^z\},\{\sigma_{e,j}^z\}} \prod_{i=1}^{N_p} \prod_{j=1}^{N_e} \langle\sigma_{p,{\cal P}_p(i)}^z|\sigma_{p,i}^z\rangle  \langle\sigma_{e,{\cal P}_e(j)}^z|\sigma_{e,j}^z\rangle \\ \times \int_{\Lambda^N} \prod_{i=1}^{N_p} \dd \textbf{R}_i \prod_{j=1}^{N_e} \dd \textbf{r}_j
\big\langle \textbf{R}_{{\cal P}_p(1)}...\textbf{R}_{{\cal P}_p(N_p)}\textbf{r}_{{\cal P}_e(1)}...\textbf{r}_{{\cal P}_e(N_e)}\big|\exp (-\beta H_{N_p,N_e})\big|\textbf{R}_1...\textbf{R}_{N_p}\textbf{r}_1...\textbf{r}_{N_e}\big\rangle
\label{quatreun}
\end{multline} 
In (\ref{quatreun}), ${\cal P}_{p,e}$ is a permutation of $(1,...,N_{p,e})$, and $\epsilon ({\cal P}_{p,e})$ is the signature $(\pm 1)$ of  ${\cal P}_{p,e}$. Notice that the spin part of the matrix elements contributes the degeneracy factor  $\sum_{\{\sigma_{p,i}^z\},\{\sigma_{e,j}^z\}}$ $\prod_{i=1}^{N_p} \prod_{j=1}^{N_e} \langle\sigma_{p,{\cal P}_p(i)}^z|\sigma_{p,i}^z\rangle \langle\sigma_{e,{\cal P}_e(j)}^z|\sigma_{e,j}^z\rangle$ which only depends on the permutations ${\cal P}_{p,e}$.

The expression (\ref{quatreun}) can now be transformed by using the Feynman-Kac path integral representation \cite{SimSchRoe} of all diagonal and off diagonal matrix elements of 
$\exp (-\beta H_{N_p,N_e})$, i.e.
\begin{multline}
\langle \textbf{x}_1'...\textbf{x}_N'|\exp (-\beta H_{N_p,N_e})| \textbf{x}_1...\textbf{x}_N \rangle
= \prod_{i=1}^N {\exp [-(\textbf{x}_i' - \textbf{x}_i)^2/(2\lambda_{\alpha_i}^2)] \over (2\pi \lambda_{\alpha_i}^2)^{3/2}}   
\int \prod_{i=1}^N {\cal D}(\bbox{\xi}_i) \\
\exp \!\Big[\!-{\beta \over 2} \sum_{i\neq j} e_{\alpha_i} e_{\alpha_j} \int_0^1 \dd s \, v(|(1-s)(\textbf{x}_i-\textbf{x}_j) + s(\textbf{x}_i'-\textbf{x}_j') + \lambda_{\alpha_i}\bbox{\xi}_i(s) - \lambda_{\alpha_j}\bbox{\xi}_j(s)|)\Big]
\label{quatredeux}
\end{multline}  
In (\ref{quatredeux}), the functional integration is carried over all Brownian bridges $\bbox{\xi}_i(s)$ subjected to the constraint $\bbox{\xi}_i(0) = \bbox{\xi}_i(1) = 0$, with the normalized Gaussian measure ${\cal D}(\bbox{\xi})$ defined by its covariance
\begin{equation}
\int {\cal D}(\bbox{\xi}) \bbox{\xi}_{\mu}(s) \bbox{\xi}_{\nu}(t) = \delta_{\mu \nu} \inf (s,t) (1 - \sup (s,t))
\label{quatretrois}
\end{equation}
In the genuine formulation introduced by Feynman, $(1 - s)\textbf{x}_i' + s\textbf{x}_i + \lambda_{\alpha_i} \bbox{\xi}_i(s)$ is the position of the particle $i$ at the dimensionless time $s$ (in units of $\beta \hbar$) along a path joining $\textbf{x}_i $ to $\textbf{x}_i'$. If we set $\textbf{x}_i' = \textbf{R}_{{\cal P}_p(i)}$ for $i=1,...,N_p$, and $\textbf{x}_i' = \textbf{r}_{{\cal P}_e(j)}$ with $j=i-N_p$ for $i=N_p+1,...,Np+N_e$, we see that this path is closed if ${\cal P}_{p}(i) = i$ or ${\cal P}_{e}(j) = j$, and is open otherwise. However, since any permutation can be decomposed into a product of cyclic permutations, all open paths associated with the particles exchanged in a given cycle can be collected in a bigger closed path. Thus, as first used by Ginibre~\cite{Gin}, $\Xi$ then becomes identical to the grand-partition function $\Xi_{loop}$ of a classical system of loops with Maxwell-Boltzmann statistics. The Boltzmann factor for loops was reorganized by Cornu~\cite{Cor1} in order to exhibit a generalized loop fugacity and two-body interactions between loops. This leads to \cite{Cor1,BryMar}
\begin{equation}
\Xi = \Xi_{loop} = \sum_{N=0}^{\infty} {1 \over N!} \int \prod_{i=1}^N {\cal D}({\cal L}_i) z({\cal L}_i) \prod_{i<j} \exp (-\beta V({\cal L}_i, {\cal L}_j)),
\label{quatrequatre}
\end{equation}
where the various symbols and quantities are defined as follows. 

First, a loop ${\cal L}$ is characterized by several degrees of freedom $\textbf{X}$, $\alpha$, $q$, $\bbox{\eta}(s)$ with $s \in [0,q]$. $\textbf{X}$ is the position of the loop, while $\alpha$ specifies the nature (electrons or protons) of the $q$ exchanged particles in the loop. The shape of the loop $\bbox{\eta}(s)$ is determined by parametrizing the collection of the $q$ trajectories followed by the $q$ exchanged particles as $\textbf{X} + \lambda_{\alpha}\bbox{\eta}(s)$ with $s \in [0,q]$. Within this parametrization, the genuine particle positions in the matrix elements are of the form $\textbf{x}^{(k)}=\textbf{X} + \lambda_{\alpha}\bbox{\eta}(k)$ with $k$ integer, $k=0,...,q-1$, and each trajectory connects $\textbf{x}^{(k)}$ to $\textbf{x}^{(k+1)}$ with $\textbf{x}^{(0)}=\textbf{x}^{(q)}=\textbf{X}$ 
(see Fig.1). The closed Brownian path $\bbox{\eta}(s)$, $\bbox{\eta}(0)=\bbox{\eta}(q)=0$, is distributed according to the normalized Gaussian measure ${\cal D}(\bbox{\eta})$ with covariance
\begin{equation}
\int {\cal D}(\bbox{\eta}) \eta_{\mu}(s) \eta_{\nu}(t) =\delta_{\mu \nu} q \inf (s/q,t/q) \big(1 - \sup (s/q,t/q)\big).
\label{quatrequatrea}
\end{equation}
\begin{figure}[h]
\myfigure
{\psfrag{X0}{$\textbf{X}$}
\psfrag{X1}{$\textbf{x}^{(1)}$}
\psfrag{X2}{$\textbf{x}^{(2)}$}
\psfrag{X3}{$\textbf{x}^{(3)}$}
\psfrag{X4}{$\textbf{x}^{(4)}$}
\psfrag{Brownian path}{Brownian path}
\psfrag{...}{$\bbox{X} + \lambda_\alpha\bbox{\eta}(s)$}
\includegraphics{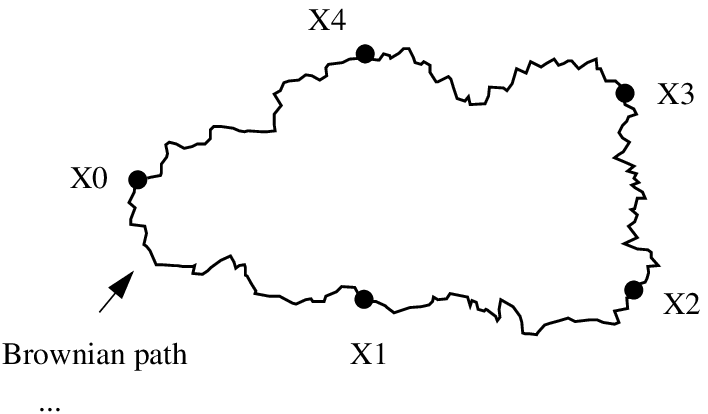}
}
{\label{fig2}
A loop ${\cal L}=(\alpha,q,\textbf{X},\bbox{\eta}(s))$ made of 5 particles.
}
{r}
\end{figure}
In deriving (\ref{quatrequatrea}), we have used that each individual trajectory is itself a Brownian path with ending points that are Gaussianly distributed. 
According to all the external and internal degrees of freedom which define the state of a loop, the phase-space measure ${\cal D}({\cal L})$ is the product of the discrete summations over $\alpha$ and $q$, the spatial integration over $\textbf{X}$ and the functional integration over $\bbox{\eta}(s)$ with the Gaussian measure ${\cal D}(\bbox{\eta})$, i.e.
\begin{equation}
\int {\cal D}({\cal L})...=\sum_{\alpha=p,e} \sum_{q=1}^{\infty} \int \dd \textbf{X} \int {\cal D}(\bbox{\eta})...
\label{quatrequatreb}
\end{equation} 
The fugacity $z({\cal L})$ reads \cite{Cor1,BryMar}

\begin{equation}
z({\cal L}) = (-1)^{q-1} {2 \over q} {z_{\alpha}^q \over 
(2\pi q\lambda_{\alpha}^2)^{3/2}} \exp (-\beta U({\cal L})),
\label{quatrecinq}
\end{equation}
where the signature $(-1)^{q-1}$ of the cyclic permutation of the $q$ exchanged particles accounts for Fermi statistics, the factor $2$ arises from spin degeneracy, and the self energy $U({\cal L})$ of the loop is 
\begin{equation}
U({\cal L}) = {e^2 \over 2} \int_0^q \dd s \int_0^q \dd t (1-\delta_{[s],[t]}) \tilde{\delta}(s-t) \,v\big(|\lambda_{\alpha}\bbox{\eta}(s) - \lambda_{\alpha}\bbox{\eta}(t)|\big)
\label{quatresix}
\end{equation}
with the Dirac comb 
\begin{equation}
\tilde{\delta}(s-t) = \sum_{n=-\infty}^{\infty} \delta (s-t-n).
\label{quatresixa}
\end{equation}
In (\ref{quatresix}), $[s]$ ($[t]$) denotes the integer part of $s$ ($t$), 
so the factor $(1-\delta_{[s],[t]})$ excludes the self energies of each individual particle. Eventually, the two-body potential $V({\cal L}_i, {\cal L}_j)$ reduces to\footnote{Notice that contrary to eq. (5.11) of \cite{BryMar}, the factor $e_{\alpha_i}  e_{\alpha_j}$ is here included in the definition of $V({\cal L}_i, {\cal L}_j)$.} 

\begin{equation}
V({\cal L}_i, {\cal L}_j) = e_{\alpha_i}  e_{\alpha_j} \int_0^q \dd s \int_0^q \dd t \, \tilde{\delta}(s-t)\, v\big(|\textbf{X}_i + \lambda_{\alpha_i}\bbox{\eta}_i(s) - \textbf{X}_j - \lambda_{\alpha_j}\bbox{\eta}_j(t)|\big).
\label{quatresept}
\end{equation}
In (\ref{quatresept}), the Dirac comb (\ref{quatresixa}) ensures that each line element of $\bbox{\eta}_i(s)$ interacts only with the corresponding line element of $\bbox{\eta}_j (t)$ taken at a time $t$ which differs from $s$ by an integer. This ``equal-time'' condition can be seen as a manifestation of the quantum nature of the particles. Consequently the potential (\ref{quatresept}) between two loops is different from the electrostatic interaction energy between uniformally charged wires with the same shapes. This difference is the source of  algebraic tails in the equilibrium correlations of quantum particles~\cite{AlaMar,CorMar,Cor2}.

According to the equivalence formula (\ref{quatrequatre}), the equilibrium quantities of the quantum gas can be inferred from that of the system of loops.
For instance, the particle densities $\rho_{\alpha}$ are related to the one-body loops density $\rho ({\cal L}_a)$ via the identity~\cite{Cor1,BryMar},

\begin{equation}
\rho_{\alpha} = \sum_{q_a=1}^{\infty} \int {\cal D}(\bbox{\eta}_a) q_a \rho ({\cal L}_a),
\label{quatresepta}
\end{equation}  
where $\alpha$ is the species index of ${\cal L}_a$.

The introduction of the gas of loops is quite useful for our purpose because the standard methods of classical statistical mechanics can be applied to this system. In particular, the equilibrium distribution functions of the loops system are formally represented by series of generalized Mayer graphs. In each graph, the usual points are now replaced by loops~\cite{Cor1}. Each loop ${\cal L}_i$ carries a statistical weight $z({\cal L}_i)$. Two loops ${\cal L}_i$ and  ${\cal L}_j$ are connected by at most one Mayer bond

\begin{equation}
f({\cal L}_i, {\cal L}_j) = \exp (-\beta V({\cal L}_i, {\cal L}_j)) - 1,
\label{quatrehuit}
\end{equation}
and the other standard topological rules \cite{HanMcd} hold. 
The contribution of a given graph is obtained by integrating over 
all degrees of freedom of each loop, except some degrees of freedom of 
the root loops that define the arguments of the considered distribution function. For instance, loop density 
$\rho ({\cal L}_a)$ reads,
\begin{equation}
\rho ({\cal L}_a) = \sum_{{\cal G}} {1 \over S({\cal G})} z({\cal L}_a) \int \prod_n {\cal D}({\cal L}_n) z({\cal L}_n) \Big[\prod f\Big]_{{\cal G}}.
\label{quatreneuf}
\end{equation}
where the sum runs over all unlabelled topologically different connected graphs ${\cal G}$ with one root point ${\cal L}_a$ and $N=0,1,...,$ internal points ${\cal L}_n$ (for $N=0$, the integral over the ${\cal L}_n$'s is replaced by $1$). The symmetry factor $S({\cal G})$ is the number of permutations of the labelled loops ${\cal L}_n$ that leave the product of bonds $[\prod f]_{{\cal G}}$ unchanged. The first graphs that appear in (\ref{quatreneuf}) are shown in Fig.2.
\begin{figure}[h]
\myfigure
{\includegraphics{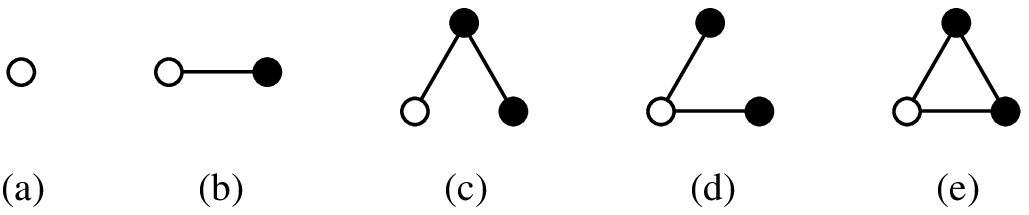}}
{	\label{fig3}
First Mayer graphs $\cal{G}$. The points stand for loops and the bonds represent Mayer factors $f=\exp(-\beta V)-1$.
}
{c}
\end{figure}

Before going further within the loop formalism, it is useful to make the connexion, at least at a formal level, between diagrammatic series (\ref{quatreneuf}) and fugacity expansions (\ref{troiscinq}), (\ref{troissix}) of particles densities $\rho_{p,e}$. First, if we sum together contributions of all graphs in (\ref{quatreneuf}) with a given number $N$ of loops, we recover the Mayer coefficient ${\cal B}_N$ 
that arises in the expansion of $\rho ({\cal L}_a)$ infered from
\begin{equation}
\rho ({\cal L}_a) =  z({\cal L}_a) {\delta \ln \Xi_{loop} \over 
\delta z({\cal L}_a)},
\label{quatreneufaa}
\end{equation}
i.e.
\begin{equation}
\rho ({\cal L}_a) = \sum_{N=1}^{\infty} { z({\cal L}_a) \over (N-1)!}
\int \prod_{i=1}^{N-1} {\cal D}({\cal L}_i) z({\cal L}_i)
{\cal B}_N ({\cal L}_a, {\cal L}_1,...,{\cal L}_{N-1}).
\label{quatreneufab}
\end{equation}
For instance, if the graphs shown in Fig.2a. and 2b. exactly provide 
${\cal B}_1 ({\cal L}_a) = 1$ and 
\begin{equation}
{\cal B}_2 ({\cal L}_a, {\cal L}_1) = 
\exp (-\beta V({\cal L}_a, {\cal L}_1)) - 1
\label{quatreneufa}
\end{equation}
respectively, the sum of three graphs with $N=2$ shown in Figs.2c, 2d and 2e, gives the third Mayer coefficient 
\begin{multline}
{\cal B}_3 ({\cal L}_a, {\cal L}_1, {\cal L}_2) = 
\exp (-\beta V({\cal L}_a, {\cal L}_1)) 
\exp (-\beta V({\cal L}_a, {\cal L}_2)) 
\exp (-\beta V({\cal L}_1, {\cal L}_2))  \\
- \exp (-\beta V({\cal L}_a, {\cal L}_1)) 
- \exp (-\beta V({\cal L}_a, {\cal L}_2))
- \exp (-\beta V({\cal L}_1, {\cal L}_2)) + 2.
\label{quatreneufb}
\end{multline}
The usual Mayer coefficients $B_{N_p,N_e}$ that arise in 
expansions (\ref{troiscinq}) and (\ref{troissix}) of $\rho_{p,e}$ 
with respect to $z_p$ and $z_e$ can be recovered by inserting 
(\ref{quatreneufab})
into the r.h.s. of (\ref{quatresepta}). After expressing integrations over 
loops degrees of freedom in terms of that of particles, we obtain each coefficient $B_{N_p,N_e}$ by collecting together all contributions with given numbers $N_p$ and $N_e$ of protons and electrons respectively. The expression of $B_{N_p,N_e}$ in terms of the suitably symmetrized traces (\ref{troisdeuxa}) then follows by applying backwards the Feynman-Kac formula (\ref{quatredeux}).

As mentionned in Section 2.2, all coefficients $B_{N_p,N_e}$ diverge. Of course, such divergencies are still present in the Mayer graphs ${\cal G}$. In fact, since the two-body loop interactions (\ref{quatresept})  are long-ranged and decay as the genuine Coulomb interactions between point charges,
\begin{equation}
V({\cal L}_i, {\cal L}_j) \sim {q_ie_{\alpha_i}  q_je_{\alpha_j} \over |\textbf{X}_i - \textbf{X}_j|} , \qquad  |\textbf{X}_i - \textbf{X}_j| \rightarrow \infty,
\label{quatreneufc}
\end{equation}
the spatial integrals over the relative distances 
$|\textbf{X}_i - \textbf{X}_j|$ between different loops do not converge. Such long-range divergencies are easily removed by taking advantage of the classical nature of the loops degrees of freedom. The corresponding method is a straightforward extension~\cite{AlaCorPer,Cor1} of chain resummations  
introduced for classical point charges \cite{MaySal,Mee,Abe}. It is briefly sketched in the next section where we give the resummed form of the low-fugacity diagrammatic series (\ref{quatreneuf}) for $\rho ({\cal L}_a)$.

\subsection{Resummed low-fugacity expansions}

The basic trick which allows us to remove the long-range divergencies 
consists in the resummation of the simple convolution chains built 
with the long-range part of the Mayer bond. The corresponding resummations
can be carried out in various ways, according to different possible decompositions of the Mayer bond. For our purpose, it is quite convenient 
to make the following choice~\cite{BalMarAla},
\begin{equation}
f({\cal L}_i, {\cal L}_j) = f_T({\cal L}_i, {\cal L}_j) 
-\beta V({\cal L}_i, {\cal L}_j) 
+ {\beta^2 \over 2} [V({\cal L}_i, {\cal L}_j)]^2,
\label{quatredix}
\end{equation}
which defines $f_T({\cal L}_i, {\cal L}_j)$. The replacement of each Mayer 
bond by its decomposition (\ref{quatredix}) provides new graphs built 
with three kinds of bonds, $f_T$, $-\beta V$ and $\beta^2 V^2/2$. 
Then, all simple convolution chains built with bonds $-\beta V$ can 
be resummed in terms of the sole screened potential

\begin{multline}
\phi ({\cal L}_i, {\cal L}_j) = V({\cal L}_i, {\cal L}_j) 
+ \sum_{n=1}^{\infty} (-1)^n \beta^n  \\ 
\times \int [\prod_{l=1}^n {\cal D}({\cal L}_l) z({\cal L}_l)] 
V({\cal L}_i, {\cal L}_1)...V({\cal L}_l, {\cal L}_{l+1})...
V({\cal L}_n, {\cal L}_j)
\label{quatreonze}
\end{multline}
Once all chain resummations have been performed, the whole set of Mayer graphs ${\cal G}$ is exactly transformed into a new set of prototype graphs ${\cal G}_P$ built with the resummed bonds,

\begin{equation}
F_c = -\beta \phi
\label{quatredouze} 
\end{equation}
and

\begin{equation}
F_R = \exp (-\beta \phi) - 1 + \beta \phi.
\label{quatretreize}
\end{equation}
The statistical weights $w({\cal L})$ of loops are either 
$z({\cal L})$ (bare loop), or $z({\cal L}) (\exp (I_R({\cal L})) - 1)$ 
(dressed loop). 
$I_R$ is the sum of all simple rings built with $-\beta V$; 
it is equal to
\begin{equation}
I_R({\cal L}) = {1 \over 2} \int {\cal D}({\cal L}_1) z({\cal L}_1)
\beta V({\cal L}, {\cal L}_1) \beta \phi ({\cal L}_1, {\cal L})
={\beta \over 2}(V - \phi)({\cal L}, {\cal L})
\label{quatrequatorze}
\end{equation}
The topological rules which define prototype graphs ${\cal G}_P$
are identical 
to those relative to Mayer graphs ${\cal G}$, except for the following 
additional exclusion rules that avoid double counting. The convolution 
of two bonds $F_c$ with an intermediate bare loop is forbidden. 
Moreover, when a bare loop is connected to the rest of the graph 
by a single bond, this bond is either $F_c$, or
\begin{equation}
F_R^{(T)} = F_R - {\beta^2 \phi^2 \over 2}. 
\label{quatrequatorzea}
\end{equation}
Eventually, the symmetry factors attached to the ${\cal G}_P$'s 
are defined as those of the ${\cal G}$'s. According to these rules, 
the whole series (\ref{quatreneuf}) is exactly transformed into
\begin{equation}
\rho ({\cal L}_a) = \sum_{{\cal G}_P} {1 \over S({\cal G}_P)} w({\cal L}_a)  
\int \prod_n {\cal D}({\cal L}_n) w({\cal L}_n) \Big[\prod F\Big]_{{\cal G}_P},
\label{quatrequinze}
\end{equation}
where the sum runs over all unlabelled topologically different 
prototype graphs ${\cal G}_P$ while $w$ and $F$ denote the various kinds 
of weights and bonds. The first graphs that appear in (\ref{quatrequinze}) are shown in Fig.3.
\begin{figure}[h]
\myfigure
{\psfrag{z_R}{$z(e^{I_R}-1)$}
\psfrag{z}{$z$}
\includegraphics{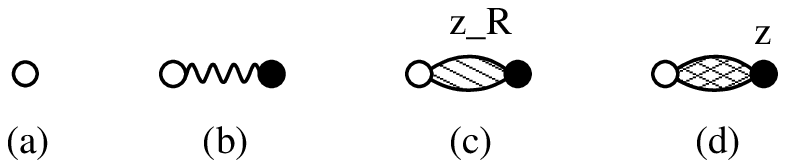}}
{	\label{fig4}
First prototype graphs $\cal{G}_P$. The possible bonds are $F_c$ (diagram b), $F_R$ (diagram c) and the ending bond $F_R^{T}$ (diagram d). If the weight of a point is not specified, it can take the two values $z$ or $z(\exp(I_R)-1)$. 
}
{c}
\end{figure}

We stress that the previous remarkably simple result follows 
from combinatorial and topological properties of Mayer graphs (see Appendix A). 
It does not depend neither on the used decomposition of the Mayer bond, 
nor on the specific form of the two-body potential. Therefore,
the technical details of these resummations are identical 
to those already displayed in previous works relative to other decompositions of $f$~\cite{Cor1,Cor3} or to relativistic interactions~\cite{AlaApp} (the corresponding proofs are simplified 
versions of the original derivation devised by Meeron~\cite{Mee} 
for classical point charges). The decomposition used in \cite{Cor1} 
follows by expressing the two-body loops potential $V$
as a sum of multipolar interactions. Then, the resummations of all 
convolution chains built with charge-charge interactions lead to diagrammatic 
series analogous to (\ref{quatrequinze}). The corresponding graphs 
have the same topological structure as the ${\cal G}_P$'s, with different weights, bonds and exclusion rules. A brief comparison between 
both (equivalent) series can be found in \cite{BalMarAla}.  

The resummed potential $\phi$ is the quantum analogue of Debye potential 
\cite{BalMarAla}. In a dilute regime, degeneracy effects for free protons 
and free electrons are weak, so the classical Debye screening length 
$\kappa^{-1} = (4 \pi \beta \sum_{\alpha} e_{\alpha}^2 \rho_{\alpha}^{id})^{-1/2} = 
(16 \pi \beta e^2 z/(2\pi \lambda^2)^{3/2})^{-1/2}$ naturally emerges. 
More precisely, at large (but not infinite)
distances $|\textbf{X}_i - \textbf{X}_j| \sim \kappa^{-1}$, 
$\phi ({\cal L}_i, {\cal L}_j)$ behaves as the classical Debye potential 
between two point charges $q_i e_{{\alpha}_i}$ and $q_j e_{{\alpha}_j}$ 
with screening length $\kappa^{-1}$,
\begin{equation}
\phi ({\cal L}_i, {\cal L}_j) \sim q_i e_{{\alpha}_i} q_j e_{{\alpha}_j}
{\exp (-\kappa |\textbf{X}_i - \textbf{X}_j|) \over 
|\textbf{X}_i - \textbf{X}_j|}.
\label{quatreseize}
\end{equation}
At larger distances, 
$|\textbf{X}_i - \textbf{X}_j| \gg \kappa^{-1}$,   
$1/|\textbf{X}_i - \textbf{X}_j|^3$-dipolar interactions 
overcome the
exponentially decaying Debye term. Out of the framework of
the loops formalism, this mechanism can be seen as an irreducible 
manifestation of quantum mechanics that ultimately leads to algebraic tails 
in the particle correlations. Eventually, at short distances 
$|\textbf{X}_i - \textbf{X}_j| \ll \kappa^{-1}$, 
$\phi ({\cal L}_i, {\cal L}_j)$ reduces to the bare potential
$V ({\cal L}_i, {\cal L}_j)$ apart from the constant 
$-q_i e_{{\alpha}_i} q_j e_{{\alpha}_j} \kappa$.  

According to the previous properties of $\phi$, both spatial and functional integrations in any prototype graph  ${\cal G}_P$ do converge \cite{BalMarAla}. 
The $1/|\textbf{X}_i - \textbf{X}_j|^3$-decay of $\phi$ removes all 
long range Coulomb divergencies. Since all Coulomb singularities at the origin are smoothed out by functional integrations over shapes of loops, 
integrability at short distances is also fulfilled.
This smearing process is nothing but the consequence of
Heisenberg's uncertainty principle which prevents a quantum
particle to stay at a fixed position (a similar 
regularization also occurs 
for Mayer graphs ${\cal G}$ built with $V$). Furthermore, 
the exponentiated structure of Boltzmann factors, which is an essential
ingredient for the description of short-range effects,
is preserved through the resummation process. Indeed, both bonds 
$F_R$ and $F_R^{(T)}$ provide factors $\exp (-\beta \phi)$ 
in the ${\cal G}_P$'s, which can be replaced by $\exp (-\beta V)$ 
in a dilute regime: this ensures that recombination can be properly 
taken into account after a suitable reorganization of 
prototype graphs (see Sections 4 and 5).

As a conclusion, resummed diagrammatic series (\ref{quatrequinze}) 
is well suited for our purpose. The resummed potential $\phi$ does account 
for classical or quantum mechanisms that prevail at either short or 
large distances. Here, it should be noticed that each prototype graph
${\cal G}_P$ involves infinite sums over numbers $q_i$ of particles contained 
in internal loops ${\cal L}_i$. Though each contribution
associated with a given set $\{q_i\}$ is finite because both functional 
and spatial integrations converge, the discrete sum over all $q_i$'s 
may diverge~\cite{BalMarAla}. Therefore, each prototype graph ${\cal G}_P$ 
must be understood as a formal infinite collection of graphs, 
with identical structures in terms of loops and different particles numbers
inside the loops.

\section{Resummed expansions in terms of particle clusters}

In this Section, series (\ref{quatrequinze}) 
will be reorganized  in terms of new diagrams made with finite clusters 
of particles. In a given 
cluster, all particles interact together and the weights are correctly
symmetrized according to Fermi statistics. The reorganization is similar
to that described in Section 3.1 for recovering the coefficients
$B_{N_p,N_e}$ from the Mayer series (\ref{quatreneuf}).

\subsection{Introduction of clusters made with fully interacting loops}
 
In each given graph ${\cal G}_P$, if we replace all bonds by their 
expressions in terms of $\phi$, products of Boltzmann factors 
$\exp (-\beta \phi)$ appear. A given product of Boltzmann factors 
then defines a cluster of interacting loops. However, such a product 
may be incomplete in the sense that it does not incorporate 
all Boltzmann factors associated with every pair of loops. 
\begin{figure}[h]
\myfigure
{\psfrag{z}{$z$}
\includegraphics{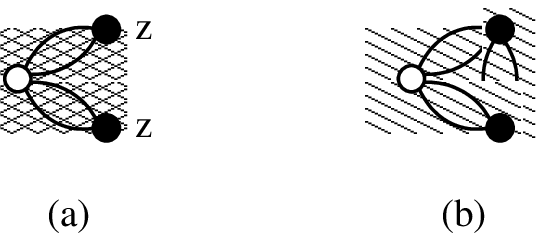}}
{	\label{fig5}
Prototype graphs showing incomplete Boltzmann factors.
}
{r}
\end{figure}
For instance, if we consider the prototype graph shown in Fig.4a, 
the product of bonds 
$\prod F = F_R^{(T)}({\cal L}_a, {\cal L}_1) 
F_R^{(T)}({\cal L}_a, {\cal L}_2)$ provides the term 
$ \exp [-\beta \phi ({\cal L}_a, {\cal L}_1)] 
\exp [-\beta \phi ({\cal L}_a, {\cal L}_2)]$. This term defines 
the set $\{ {\cal L}_a, {\cal L}_1, {\cal L}_2 \}$ of interacting loops, 
and it is incomplete since the Boltzmann factor 
$ \exp [-\beta \phi ({\cal L}_1, {\cal L}_2)]$ associated 
with pair $({\cal L}_1, {\cal L}_2)$ is missing. Since
all mutual interactions between particles that belong to loops
$\{ {\cal L}_a, {\cal L}_1, {\cal L}_2 \}$ are not present, the considered graph
provides unphysical contributions. Of course, such contributions must 
be cancelled out by other similar contributions arising in other graphs. 
In fact, if we add the contribution from graph shown in Fig.4b 
to that 
from graph in Fig.4a, we see that the previous incomplete product  
$ \exp [-\beta \phi ({\cal L}_a, {\cal L}_1)] 
\exp [-\beta \phi ({\cal L}_a, {\cal L}_2)]$ does disappear. Here,
we first proceed to a systematic reorganization of the whole diagrammatic 
series (\ref{quatrequinze}) where all incomplete products of Boltzmann 
factors are removed.    

According to Section 3.2, graphs ${\cal G}_P$ are
identical to Mayer graphs of a system with $\phi$ in place of $V$ and 
$z\exp(I_R)$ in place of $z$, except 
for some exclusion rules that avoid double counting. In the absence of such rules, the (finite) sum of graphs ${\cal G}_P$ with $N$ loops 
would provide the Mayer coefficient $ {\cal B}_{\phi,N}$, similarly 
to the derivation of ${\cal B}_N$ from the ${\cal G}$'s described in 
Section 3.1. Since all Boltzmann factors are complete in ${\cal B}_{\phi,N}$, 
the required reorganization would then be achieved. Here, the difficulty entirely arises from the exclusion rules which prevent to merely 
perform the previous sum. Our method consists in introducing parts of 
graphs made with $M$ loops which are not affected by the exclusion rules, 
and for which summations over all possible internal structures provide 
the coefficient ${\cal B}_{\phi,M}$. The remaining parts are built with either convolutions of bonds $F_c$, or bonds $F_c^2/2$, so all Boltzmann factors 
are eventually incorporated in the ${\cal B}_{\phi,M}$'s.

The first step of the method is based on the introduction of new graphs 
$ {\cal G}_P^{\ast}$ defined as follows. In each graph ${\cal G}_P$, we decompose all ending bonds 
$F_R^{(T)}$ as $F_R$ plus $-\beta^2 \phi^2/2 = -F_c^2/2$. 
At the same time, we rewrite all dressed weights 
$z({\cal L}) (\exp (I_R({\cal L})) - 1)$ of intermediate loops in 
convolutions $F_c \star F_c$ as  $z({\cal L})\exp (I_R({\cal L}))$ plus
$-z({\cal L})$. The series (\ref{quatrequinze}) then becomes
\begin{equation}
\rho ({\cal L}_a) = \sum_{{\cal G}_P^{\ast}} {1 \over S({\cal G}_P^{\ast})}  
 w^{\ast}({\cal L}_a) \int \prod_n {\cal D}({\cal L}_n) w^{\ast}({\cal L}_n) \Big[\prod F^{\ast}\Big]_{{\cal G}_P^{\ast}}.
\label{cinqun}
\end{equation}
The modified prototype graphs $ {\cal G}_P^{\ast}$ are built with 
three kinds of bonds $F^{\ast}$ that are $F_c$, $F_R$ or $F_c^2/2$, 
and four kinds of weights $w^{\ast}$ that are $z$, $z (\exp(I_R)-1)$, $z \exp(I_R)$ or $z^{\ast} = -z$. They have the same topological structure as 
the genuine Mayer graphs except for the following rules. The
intermediate loop in a convolution $F_c \star F_c$ carries a weight
which is either $z \exp(I_R)$ or $z^{\ast}$. Each bond $F_c^2/2$ is an ending
bond that connects a single loop with weight $z^{\ast}$ to the 
rest of the graph. All other loops (including the root loop ${\cal L}_a$) 
carry weights that are either    
$z$ or $z(\exp(I_R)-1)$. The sum in (\ref{cinqun}) runs over all non topologically equivalent graphs. The symmetry factor 
$ S({\cal G}_P^{\ast})$ is defined as usual by virtue of the 
combinatorial properties derived in Appendix~A. The first graphs  
$ {\cal G}_P^{\ast}$ are shown in Fig.5.
\begin{figure}[h]
\myfigure
{\psfrag{z_R}{$z(e^{I_R}-1)$}
\psfrag{z}{$z$}
\psfrag{z*}{$z^*$}
\includegraphics{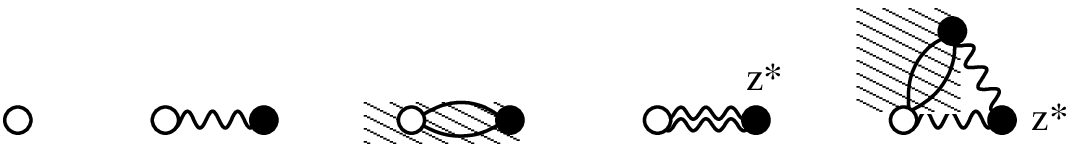}}
{	\label{fig6}
First graphs $\cal{G}_P^*$. The ending bond $F_c^2/2$ is drawn as a double wavy line. As before, the points with no weight specified can take the two values $z$ or $z\exp(I_R)-1$.
}
{c}
\end{figure}

In each graph  $ {\cal G}_P^{\ast}$, 
we can make a partition of the ensemble of loops as follows. On one hand,
there are loops with weights $z^{\ast}$, and on
another hand, there remains all other loops which can be distributed 
into clusters. These clusters are defined as the 
remaining connected subsets when 
the loops with weights $z^{\ast}$ are removed. 
In each cluster ${\cal C}$, bonds
are either $F_c$ or $F_R$, and weights are either $z$ or 
$z (\exp(I_R)-1)$ except for the intermediate loops in convolutions 
$F_c \star F_c$ which carry weights $z \exp(I_R)$. A given loop in a cluster
may be connected to another loop, in the same cluster or in another one, 
by  convolution chains $F_c \star F_c ... \star F_c$ where all 
intermediate loops carry weights $z^{\ast}$; or  
a similar ring convolution with at least two intermediate 
loops can be attached to this loop. It can  be also connected to ending loops 
with weights $z^{\ast}$ by single bonds $F_c^2/2$. Several examples
of the cluster structure of the graphs $ {\cal G}_P^{\ast}$ are shown 
in Fig.6.
\begin{figure}[h]
\myfigure
{\psfrag{z_R}{$z(e^{I_R}-1)$}
\psfrag{z}{$z$}
\psfrag{z*}{$z^*$}
\includegraphics{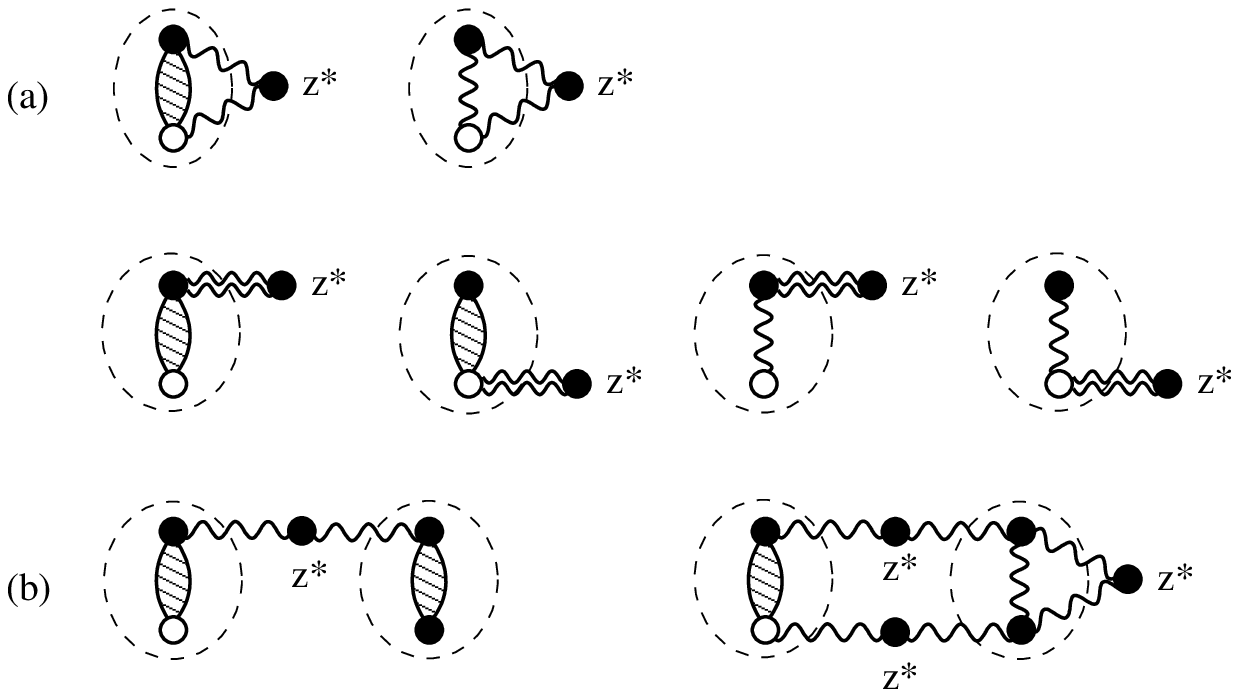}}
{	\label{fig7}
Cluster structures in the graphs ${\cal G^*_P}$. (a) graphs belonging to the same class and leading to a graph ${\cal  G_C}$ when added together. (b) two cluster structures belonging to different classes.
}
{c}
\end{figure}

Now we proceed to a classification of graphs ${\cal G}_P^{\ast}$ 
according to their topological structure in terms of the previous clusters 
and connections. All graphs in a given
class have the same respective numbers of clusters 
and loops inside each cluster. Moreover, two loops connected 
by convolution chains $F_c \star F_c ...\star F_c$ (with 
precribed numbers of intermediate loops) belong to the same respective 
pairs of clusters, or to the same respective clusters; similarly, 
rings $F_c \star F_c ...\star F_c$ (with 
precribed numbers of intermediate loops) and ending bonds 
$F_c^2/2$ are attached to loops which belong to the same 
respective clusters. Various classes of graphs ${\cal G}_P^{\ast}$ 
are shown in Fig.6. In each class, the chains can be attached to all the 
different loops inside a given cluster. Remember that, 
in such a cluster, the loops can carry any weight except $z^{\ast}$,
and the internal bonds are $F_c$ or $F_R$. Moreover, the connecting 
structure of convolution chains between different clusters are identical. 
Also, the rings or ending bonds $F_c^2/2$ attached to single clusters are identical.

It turns out that the sum of all graphs ${\cal G}_P^{\ast}$ in a given
class can be merely rewritten as a single graph ${\cal G}_{{\cal C}}$. 
Indeed, by virtue of combinatorial properties of Mayer graphs (see Appendix A), the various 
bonds and weights inside each respective clusters, as well as the various 
chains (rings) $F_c \star F_c ...\star F_c$ and ending bonds $F_c^2/2!$, can be 
added together in a simultaneous and independent way. Inside each
cluster, the sum of the two possible internal bonds $F_c$ and $F_R$ provides
the Mayer bond $f_{\phi} = \exp (-\beta \phi) - 1$ associated with $\phi$, 
while the corresponding sum of weights reduces to 
$ z_{\phi} =  z\exp (I_R)$. The further summation of all possible 
internal structures is analogous to that performed when relating usual 
Mayer graphs to Mayer coefficients (see Section 3.1). 
The corresponding final statistical weight $W_{\phi}({\cal C})$ for a cluster  
${\cal C} = {{\cal L}_1,..., {\cal L}_N}$ made with $N$ loops reads
\begin{equation}
W_{\phi}({\cal C}) = {\prod_{{\cal L} \in {\cal C}} z_{\phi}({\cal L}) 
\over N!} 
{\cal B}_{\phi}({\cal C})
\label{cinqdeux}
\end{equation}
where ${\cal B}_{\phi}({\cal C}) = {\cal B}_{\phi,N}$ is the Mayer coefficient for the $N$ loops inside
${\cal C}$ and interacting \textsl{via} $\phi$ (if the root loop ${\cal L}_a$ 
belongs to ${\cal C}$, $N!$ is replaced by $(N-1)!$). The expressions of 
${\cal B}_{\phi,N}$ are identical to that of ${\cal B}_N$ with $\phi$ 
in place of $V$ (for instance see formulas (\ref{quatreneufa}) and (\ref{quatreneufb}) for $N=2$ and $N=3$ respectively). The previous 
simultaneous resummations inside each cluster also amount to sum all bonds 
$F_c$ that connect loops belonging to a given cluster ${\cal C}$ 
to a loop ${\cal L}_i$
with weight $z^{\ast}$. Since $F_c = -\beta \phi$, this gives rise to the 
total potential $\Phi ({\cal C}, {\cal L}_i)$ between ${\cal C}$ 
and ${\cal L}_i$,
\begin{equation}
\Phi ({\cal C}, {\cal L}_i) = \sum_{{\cal L} \in {\cal C}} 
\phi ({\cal L}, {\cal L}_i),
\label{cinqtrois}
\end{equation}
while the corresponding bond ${\cal F}_c$ is 
\begin{equation}
{\cal F}_c = -\beta \Phi .
\label{cinqtroisa}
\end{equation}
Of course, all these summations do not change the topological stucture
of the chains, rings and ending bonds attached to the various clusters.
Eventually, the series (\ref{cinqun}) is exactly transformed into
\begin{equation}
\rho ({\cal L}_a) = \sum_{{\cal G}_{{\cal C}}} 
{1 \over S({\cal G}_{{\cal C}})} \int \prod_i {\cal D}({\cal C}_i)
W_{\phi}({\cal C}_i)   
\int \prod_n {\cal D}({\cal L}_n) z^{\ast}({\cal L}_n) \Big[\prod {\cal F}_c\Big]_{{\cal G}_{{\cal C}}}
\Big[\prod {{\cal F}_c^2 \over 2!}\Big]_{{\cal G}_{{\cal C}}}.
\label{cinqquatre}
\end{equation}
The integration ${\cal D}({\cal C}_i)$ over the phase space available 
to cluster 
${\cal C}_i$ reduces to integrations over degrees of freedom 
of internal loops inside ${\cal C}_i$,
\begin{equation}
{\cal D}({\cal C}_i) = \prod_{{\cal L} \in {\cal C}_i} {\cal D}({\cal L})   
\label{cinqcinq}
\end{equation}
The root loop ${\cal L}_a$ necessarily belongs to one cluster, i.e. 
${\cal C}_a = {\cal C}_0$, for which the product in
(\ref{cinqcinq}) only runs over all the other internal loops in ${\cal C}_a$. 
Each bond ${\cal F}_c = -\beta \Phi$ connects either, one cluster to one
loop with weight $z^{\ast}$, or two loops with weights $z^{\ast}$. All
these bonds are distributed into convolution chains
that connect two clusters, or into rings attached to a given cluster. 
Each bond ${\cal F}_c^2/2!$ connects one cluster to one ending loop
with weight $z^{\ast}$. The 
loops ${\cal L}_n$ with weights $z^{\ast}({\cal L}_n)$ are intermediate
loops in the previous chains and rings, or ending loops 
(there are at least two intermediate loops in 
each ring). Every graph ${\cal G}_{{\cal C}}$ is connected. 
Two clusters may be connected by an arbitrary number of chains, while an arbitrary number of rings may be attached to each cluster. The symmetry
factor $S({\cal G}_{{\cal C}})$ is defined as the number of permutations
of clusters $\{ {\cal C}_i, i \neq 0 \}$ (except the
root cluster ${\cal C}_a$) and of loops $\{ {\cal L}_n \}$ with weights
$z^{\ast}$, that leave the product of bonds 
$[\prod {\cal F}_c]_{{\cal G}_{{\cal C}}} 
[\prod {\cal F}_c^2/2!]_{{\cal G}_{{\cal C}}}$ unchanged. 
The considered
permutations separately act on loops $\{ {\cal L}_n \}$ with weights
$z^{\ast}$ on one hand, and on subsets of clusters with identical numbers of
internal loops on another hand; permutations of internal loops inside 
each cluster are already accounted for by the $1/N!$ factor in the definition
(\ref{cinqdeux}) of the cluster weight $W_{\phi}({\cal C})$. 
Eventually, the sum in 
(\ref{cinqquatre}) runs over all topologically different graphs
${\cal G}_{{\cal C}}$. The first graphs in (\ref{cinqquatre}) are shown
in Fig.7.
\begin{figure}[h]
\myfigure
{\psfrag{z*}{$z^*$}
\includegraphics{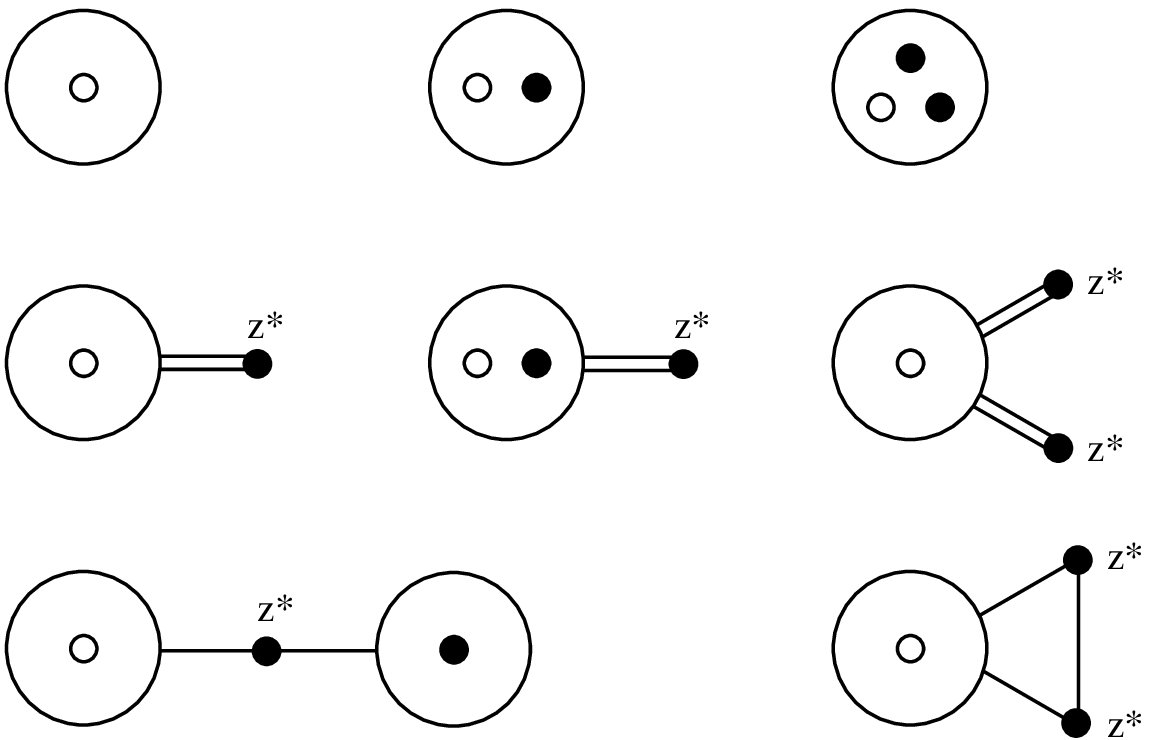}}
{\label{fig8}
First graphs ${\cal G}_{\CC}$, corresponding to the summation of all prototype graphs made of one, two and three loops. The second graph in the second row results from the summation of the graphs (a) of Fig.6. The bond ${\cal F}_c(\CC,\LL)=-\beta\Phi(\CC,\LL)$ between a cluster $\CC$ and a loop $\LL$ of weight $z^*$ is drawn as a simple line, while the bond ${\cal F}^2_c/2!$ is drawn as a double line. The statistical weight of a cluster is $W_\phi({\cal C})$.}
{c}
\end{figure}

In the diagrammatic series (\ref{cinqquatre}), the sum of all graphs 
${\cal G}_{{\cal C}}$ with a single root cluster  ${\cal C}_a$ and 
no attached rings, is nothing but
the density of a system of loops with potential $\phi$ and fugacity
$z_{\phi}$: the corresponding expression is identical to 
(\ref{quatreneufab}) with $\phi$ in place of $V$ and $z_{\phi}$ 
in place of $z$. All the other graphs, with at least 
two clusters or one ring
avoid double counting: they result from the exclusion rules for
prototype graphs ${\cal G}_P$ 
generated by the chain resummations where $V$ is replaced by $\phi$.

As required, each graph ${\cal G}_{{\cal C}}$ no longer involves incomplete 
products of Boltzmann factors $\exp (-\beta \phi)$. Indeed, all Boltzmann 
factors $\exp (-\beta \phi)$ are reorganized into Mayer coefficients 
${\cal B}_{\phi}({\cal C})$ which define the weights 
$W_{\phi}({\cal C})$ of the clusters. As shown by their construction from 
(\ref{quatreneufaa}), the ${\cal B}_N$'s involve only complete products
of factors  $\exp (-\beta V)$ associated with subsets of interacting loops 
(each product for $L$ loops arises from the $L$-th order term in the
expansion of $\Xi_{loop}$ with respect to $z$). Of course, this property 
is also satisfied by the ${\cal B}_{\phi,N}$'s with $\phi$ in place of $V$.

\subsection{Introduction of truncated Mayer coefficients}

Similarly to the contribution of any prototype graph ${\cal G}_P$ 
obtained by fixing the numbers of particles inside loops, 
the corresponding contributions of every graph ${\cal G}_{{\cal C}}$ 
are also well behaved by virtue of the sufficiently fast decay 
of the resummed potential $\phi$. In a dilute regime, at finite distances,
$\phi$ tends to $V$, so ${\cal B}_{\phi,N}$ can be merely replaced by 
${\cal B}_N$ in weights $W_{\phi}({\cal C})$. However, this replacement is
no longer legitimate for large distances between the loops inside
cluster  ${\cal C}$: it is crucial to keep 
screening effects embodied in $\phi$ for ensuring integrability. In order 
to circumvent this difficulty, we introduce truncated Mayer coefficients
${\cal B}_{\phi,N}^T$ such that their analogues ${\cal B}_N^T$ with $V$ 
in place of $\phi$ remain integrable at large relative distances. 
Of course, there are several ways of defining truncated coefficients 
by appropriate substraction of terms which are not integrable when 
$\phi$ is replaced by $V$. The definition used below is aimed at preserving 
the topological structure of Mayer graphs. Indeed, within this definition, 
we find that the resulting diagrammatic series has a standard Mayer form 
in terms of the ${\cal B}_N^T$'s (see (\ref{cinqseize})).

Of course, no truncation is necessary for ${\cal B}_{\phi,1} = 1$ since
there is only one loop in the associated cluster. So, we first consider  
${\cal B}_{\phi,2}$ that is given by formula (\ref{quatreneufa}) with
$\phi$ in place of $V$. We define ${\cal B}_{\phi,2}^T$ as
\begin{equation}
{\cal B}_{\phi,2}^T = {\cal B}_{\phi,2} + \beta \phi 
-{\beta^2 \phi^2 \over 2!} + {\beta^3 \phi^3 \over 3!}.
\label{cinqsix}
\end{equation}
By construction, ${\cal B}_{\phi,2}^T$ behaves as $\beta^4 \phi^4/4!$ 
at large distances. Therefore, ${\cal B}_2^T({\cal L}_1, {\cal L}_2)$ decays as 
$1/|\textbf{X}_1 - \textbf{X}_2|^4$ when 
$|\textbf{X}_1 - \textbf{X}_2| \rightarrow \infty$, so ${\cal B}_2^T$ 
indeed is integrable over the whole space. Notice that integrability 
at zero separation results from the smearing process already evocated 
in Section 4.2.

In order to define ${\cal B}_{\phi,3}^T$, we replace each Boltzmann factor
$\exp (-\beta \phi)$ into ${\cal B}_{\phi,3}$ by
\begin{equation}
\exp (-\beta \phi) = 1 - \beta \phi + {\beta^2 \phi^2 \over 2!}
-{\beta^3 \phi^3 \over 3!} + {\cal B}_{\phi,2}^T.
\label{cinqsept}
\end{equation}
The resulting expression for 
${\cal B}_{\phi,3}({\cal L}_1, {\cal L}_2, {\cal L}_3)$ is a (finite) sum
of terms that is nothing but the sum of Mayer diagrams where the 
Mayer bond $f_{\phi}$ has been decomposed according to (\ref{cinqsept}). 
This provides various bonds  
$g$ between two loops which can be either $F_c$, $F_c^2/2!$, $F_c^3/3!$ or 
${\cal B}_{\phi,2}^T$. Then, 
${\cal B}_{\phi,3}({\cal L}_1, {\cal L}_2, {\cal L}_3)$ can be rewritten as
\begin{equation}
{\cal B}_{\phi,3}({\cal L}_1, {\cal L}_2, {\cal L}_3) = 
\sum_{D_3} \Big[\prod g\Big]_{D_3}.
\label{cinqhuit}
\end{equation}
In (\ref{cinqhuit}), $D_3$ is a connected labelled diagram 
made with the three loops 
$({\cal L}_1, {\cal L}_2, {\cal L}_3)$, such that two loops are connected 
by at most one bond $g$. Moreover, $[\prod g]_{D_3}$ denotes the product 
of bonds in 
$D_3$, and the sum is taken over all the different diagrams (for which 
the products  $[\prod g]_{D_3}$ are different); of course no integration
is performed in $D_3$ over loops degrees of freedom. The truncated coefficient
${\cal B}_{\phi,3}^T$ is defined as the contribution to (\ref{cinqhuit})
of all diagrams which remain integrable with respect to all relative 
distances when $\phi$ is replaced by $V$. For diagrams which are built
with bonds $F_c$, $F_c^2/2!$ or $F_c^3/3!$, the integrability condition 
is fulfilled by a simple power counting. Indeed, since $V$ decays as 
$1/|\textbf{X}_i - \textbf{X}_j|$, the sixth-dimensional integration 
over the two relative distances converges if the overall power of
bonds $F_c$ is larger than, or equal to $7$. This gives 
\begin{multline}
{\cal B}_{\phi,3}^T = 
{\cal B}_{\phi,2}^T {\cal B}_{\phi,2}^T {\cal B}_{\phi,2}^T 
+ \sum {\cal B}_{\phi,2}^T{\cal B}_{\phi,2}^T \\
+ \sum {\cal B}_{\phi,2}^T{\cal B}_{\phi,2}^T F_c 
+ \sum {\cal B}_{\phi,2}^T{\cal B}_{\phi,2}^T {F_c^2 \over 2!} 
+ \sum {\cal B}_{\phi,2}^T{\cal B}_{\phi,2}^T {F_c^3 \over 3!} \\
+ \sum {\cal B}_{\phi,2}^T {F_c^2 \over 2!} {F_c^2 \over 2!}
+ \sum {\cal B}_{\phi,2}^T{F_c^2 \over 2!} {F_c^3 \over 3!}  
+ \sum {\cal B}_{\phi,2}^T{F_c^3 \over 3!} {F_c^3 \over 3!} \\ 
+ \sum F_c {F_c^3 \over 3!} {F_c^3 \over 3!} 
+ \sum {F_c^2 \over 2!}{F_c^2 \over 2!} {F_c^3 \over 3!}
+ \sum {F_c^2 \over 2!}{F_c^3 \over 3!} {F_c^3 \over 3!}
+ {F_c^3 \over 3!} {F_c^3 \over 3!} {F_c^3 \over 3!},
\label{cinqneuf}
\end{multline}
where the sums run over all the (labelled) diagrams $D_3$ with the 
considered products of bonds ${\cal B}_{\phi,2}^T$, $F_c$, $F_c^2/2!$
and $F_c^3/3!$. 
According to this definition, we rewrite (\ref{cinqhuit}) as ${\cal B}_{\phi,3}^T$ plus the remaining terms that are not integrable when $\phi$ is replaced  by $V$:
\begin{multline}
{\cal B}_{\phi,3} = {\cal B}_{\phi,3}^T 
+ \sum {\cal B}_{\phi,2}^T F_c 
+ \sum {\cal B}_{\phi,2}^T  F_c F_c
+ \sum {\cal B}_{\phi,2}^T {F_c^2 \over 2!} 
+ \sum {\cal B}_{\phi,2}^T F_c {F_c^2 \over 2!} \\
+ \sum {\cal B}_{\phi,2}^T {F_c^3 \over 3!} 
+ \sum F_c F_c 
+ \sum F_c {F_c^2 \over 2!} 
+ \sum F_c {F_c^3 \over 3!}
+ \sum {F_c^2 \over 2!} {F_c^2 \over 2!} \\                                                                                                            + \sum {F_c^2 \over 2!} {F_c^3 \over 3!}                                                                                                             + \sum {F_c^3 \over 3!} {F_c^3 \over 3!}
+ F_c F_c F_c
+ \sum F_c F_c {F_c^2 \over 2!}
+ \sum F_c F_c {F_c^3 \over 3!} \\
+ \sum F_c {F_c^2 \over 2!} {F_c^3 \over 3!} 
+ \sum F_c {F_c^2 \over 2!} {F_c^2 \over 2!}
+ {F_c^2 \over 2!} {F_c^2 \over 2!} {F_c^2 \over 2!}.
\label{cinqdix}
\end{multline}
The sums have the same meaning as in (\ref{cinqneuf}) in terms of diagrams $D_3$.

Now we introduce new diagrams $D_{{\cal C}}$ by identifying 
clusters made with one or two loops in the diagrams $D_3$: 
two loops are in the same cluster if they are connected by a bond 
${\cal B}_{\phi,2}^T$. In these diagrams $D_{{\cal C}}$, 
the loops of $D_3$ are replaced by the corresponding clusters. 
The weights of clusters with one
and two loops are ${\cal B}_{\phi,1} = 1$ and ${\cal B}_{\phi,2}^T$ respectively. Two clusters are linked by at most one of the three
possible bonds ${\cal F}_c$, ${\cal F}_c^2/2!$ or ${\cal F}_c^3/3!$, where
${\cal F}_c$ is related to the total interaction potential $\Phi$ between 
the clusters, i.e. ${\cal F}_c = - \beta \Phi$. The total potential $\Phi$ 
between a cluster with two loops and a cluster with one loop
is given by (\ref{cinqtrois}), while it obviously reduces to $\phi$ if
two clusters are made with single loops. The 
bonds between a two-loops cluster and a one-loop cluster, 
result from summing together the contributions of all
diagrams $D_3$ where the two given clusters are linked by
bonds $F_c$ for ${\cal F}_c$, by bonds $F_c F_c$ or $F_c^2/2!$ for 
${\cal F}_c^2/2!$, by bonds $F_c F_c^2/2!$ or $F_c^3/3!$ for ${\cal F}_c^3/3!$. 
Similarly to (\ref{cinqdix}), the resulting expression for 
${\cal B}_{\phi,3}$ can be detailed as   
\begin{multline}
{\cal B}_{\phi,3} = {\cal B}_{\phi,3}^T 
+ \sum {\cal B}_{\phi,2}^T {\cal F}_c 
+ \sum {\cal B}_{\phi,2}^T {{\cal F}_c^2 \over 2!} 
+ \sum {\cal B}_{\phi,2}^T {{\cal F}_c^3 \over 3!} \\
+ \sum {\cal F}_c {\cal F}_c 
+ \sum {\cal F}_c {{\cal F}_c^2 \over 2!} 
+ \sum {\cal F}_c {{\cal F}_c^3 \over 3!}
+ \sum {{\cal F}_c^2 \over 2!} {{\cal F}_c^2 \over 2!} \\                                                                                                           + \sum {{\cal F}_c^2 \over 2!} {{\cal F}_c^3 \over 3!}                                                                                                             + \sum {{\cal F}_c^3 \over 3!} {{\cal F}_c^3 \over 3!}
+ {\cal F}_c {\cal F}_c {\cal F}_c
+ \sum {\cal F}_c {\cal F}_c {{\cal F}_c^2 \over 2!}
+ \sum {\cal F}_c {\cal F}_c {{\cal F}_c^3 \over 3!} \\
+ \sum {\cal F}_c {{\cal F}_c^2 \over 2!} {{\cal F}_c^3 \over 3!} 
+ \sum {\cal F}_c {{\cal F}_c^2 \over 2!} {{\cal F}_c^2 \over 2!}
+ {{\cal F}_c^2 \over 2!} {{\cal F}_c^2 \over 2!} {{\cal F}_c^2 \over 2!},   
\label{cinqonze}
\end{multline} 
where each sum runs over all labelled diagrams $D_{{\cal C}}$ 
(loops inside a cluster are labelled) with the
considered product of weights and bonds. We stress that 
definition (\ref{cinqneuf}) allows to introduce clusters which are 
connected by single bonds ${\cal F}_c$, ${\cal F}_c^2/2!$ or 
${\cal F}_c^3/3!$. If ${\cal B}_{\phi,3}^T$ had been defined as the sum 
of diagrams with bonds ${\cal B}_{\phi,2}^T$ only (the first two terms 
of (\ref{cinqneuf})), left-over terms in (\ref{cinqneuf}) like 
$\sum {\cal B}_{\phi,2}^T{\cal B}_{\phi,2}^T F_c$ or
$\sum {\cal B}_{\phi,2}^T F_c^2/2! F_c^2/2!$, could not 
have been rewritten in terms of clusters connected by the sole links 
${\cal F}_c$, ${\cal F}_c^2/2!$ or ${\cal F}_c^3/3!$.

The construction of the general truncated coefficient ${\cal B}_{\phi,N}^T$
follows from a procedure similar to that introduced above for
${\cal B}_{\phi,3}^T$. Each Boltzmann factor is replaced by
(\ref{cinqsept}) into ${\cal B}_{\phi,N}$, with the result
\begin{equation}
{\cal B}_{\phi,N} = \sum_{D_N} \Big[\prod g\Big]_{D_N}
\label{cinqdouze}
\end{equation}
where diagrams $D_N$ are defined with the same rules as the $D_3$'s.
We define  ${\cal B}_{\phi,N}^T$ as the contribution to (\ref{cinqdouze}) 
of all diagrams $D_N$ that remain integrable with respect to all 
relative distances when $\phi$ is replaced by $V$. 
For diagrams built
with bonds $F_c$, $F_c^2/2!$ or $F_c^3/3!$, the simple power-counting rule
used for ${\cal B}_{\phi,3}$ implies that the overall sum of powers of 
bonds $F_c$ is strictly larger than $(3N-3)$. However, for $N \geq 4$, this rule 
is necessary but not sufficient. For instance, the diagram shown in Fig.8
fulfills the counting rule, but it is not integrable (when $\phi$ is replaced 
by $V$) with respect to all separations.
In fact, the determination of the integrable diagrams 
requires a closer analysis of their topological structure.
\myfigure{
\psfrag{l1}{$\LL_1$}
\psfrag{l2}{$\LL_2$}
\psfrag{l3}{$\LL_3$}
\psfrag{l4}{$\LL_4$}
\includegraphics{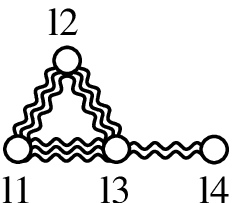}
\raisebox{-7mm}{\rule{0mm}{15mm}}}{	
\label{fig8a}
A diagram $D_4$ that fulfills the power counting rule but that is not integrable with respect to the relative distance $|\textbf{R}_4-\textbf{R}_3|$ when $\phi$ is replaced by $V$. This diagram does therefore not appear in the definition of ${\cal B}_{\phi,4}^T$. It rather belongs to the class of graphs characterized by a $F_c^2$-link between the two clusters of loops $\{\LL_1,\LL_2,\LL_3\}$ and $\{\LL_4\}$.}{r}

\noindent In each of the non-integrable diagrams, we define clusters made with $M$ 
loops ($1 \leq M < N$), for which integrals over relative distances 
between loops inside a given cluster do converge when $\phi$ is replaced 
by $V$. By construction, these clusters realize a partition 
of the ensemble of $N$ loops of the considered diagram $D_N$. 
As a result of this procedure, two 
connected clusters are linked by either one bond $F_c$  
($F_c$-link), two bonds $F_c$ or one bond $F_c^2/2!$ 
($F_c^2$-link), three bonds $F_c$, or
one bond $F_c$ and one bond $F_c^2/2!$, or one bond 
$F_c^3/3!$ ($F_c^3$-link). A diagram $D_{{\cal C}}$ is defined 
as the sum of all
diagrams $D_N$ with identical partitions into clusters and identical 
structures of $F_c^n$-links ($1 \leq n \leq 3$) between clusters. 
For a given cluster made with $M$ loops, the summation of all possible
internal structures of bonds $g$ provides, by definition, the truncated
coefficient ${\cal B}_{\phi,M}^T$ (${\cal B}_{\phi,1} = 1$ if $M=1$). 
The bond ${\cal F}_{\phi}$ 
between two connected clusters, which results from  
the sum of all possible $F_c^n$-links, reduces to 
${\cal F}_c$ for $n=1$, ${\cal F}_c^2/2!$ for $n=2$, 
or ${\cal F}_c^3/3!$ for $n=3$.   
Now, ${\cal F}_c = -\beta \Phi$, where $\Phi$ is the total interaction 
potential between two clusters, i.e.
\begin{equation}
\Phi ({\cal C}_i, {\cal C}_j) = \sum_{{\cal L}_k \in {\cal C}_i} 
\sum_{{\cal L}_l \in {\cal C}_j} \phi ({\cal L}_k, {\cal L}_l),
\label{cinqdouzea}.
\end{equation}
The detail of the summation leading to ${\cal F}_c^2/2!$ is illustrated 
in Fig.9.
\myfigure
{\psfrag{1}{$1$}
\psfrag{2}{$2$}
\psfrag{3}{$3$}
\psfrag{4}{$4$}
\includegraphics{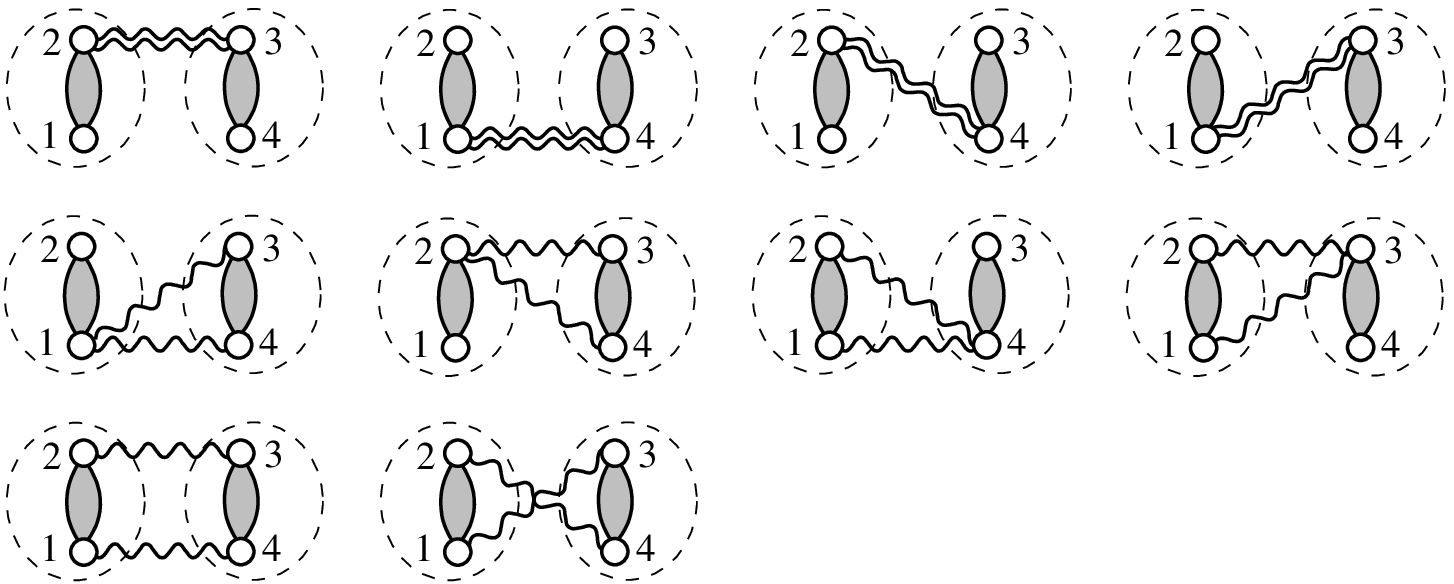}}
{\label{fig8b}
Diagrams $D_4$ belonging to the same class characterized by a $F_c^2$-link between the two clusters of loops $\{\LL_1,\LL_2\}$ and $\{\LL_3,\LL_4\}$ (the bond ${\cal B}_{\phi,2}^T$ is drawn as a gray link). Their sum leads to the $\cal D_C$ diagram ${\cal B}_{\phi,2}^T(\LL_1,\LL_2) {\cal B}_{\phi,2}^T(\LL_3,\LL_4)\FF_c^2/2!$ in~\eqref{cinqtreize}.}
{c}

\noindent Therefore, similarly to formula (\ref{cinqonze}) for $N=3$, 
we can express ${\cal B}_{\phi,N}$ in terms of truncated coefficients as \begin{equation}
{\cal B}_{\phi,N} = {\cal B}_{\phi,N}^T + 
\sum_{D_{\cal C}} \Big[\prod {\cal B}_{\phi,M}^T\Big]_{D_{\cal C}}  
\Big[\prod {\cal F}_{\phi}\Big]_{D_{\cal C}}.
\label{cinqtreize}
\end{equation}
In (\ref{cinqtreize}), the sum runs over all labelled and connected diagrams $D_{{\cal C}}$ with at least two clusters (all $M$'s are
strictly smaller than $N$), while $[\prod {\cal B}_{\phi,M}^T]_{D_{\cal C}}$  
and $[\prod {\cal F}]_{D_{\cal C}}$ denote, respectively, 
the product of weights and the product of bonds for a given diagram 
$D_{{\cal C}}$. Moreover, this sum is restricted to the diagrams   
which become not integrable over all relative distances between
clusters when $\phi$ is replaced by $V$. For instance, in the explicit 
expression (\ref{cinqonze}) of (\ref{cinqtreize}) for $N=3$, the 
diagrams $D_{{\cal C}}$ with three one-loop clusters and products of bonds 
${\cal F}_c {\cal F}_c^3/3! {\cal F}_c^3/3!$, 
${\cal F}_c^2/2! {\cal F}_c^2/2! {\cal F}_c^3/3!$,
${\cal F}_c^2/2! {\cal F}_c^3/3! {\cal F}_c^3/3!$, 
or ${\cal F}_c^3/3! {\cal F}_c^3/3! {\cal F}_c^3/3!$, do not
appear. When $N$ increases, a detailed account of the forbidden 
diagrams $D_{{\cal C}}$ becomes rapidly cumbersome, like the explicit
evaluation of the truncated coefficient ${\cal B}_{\phi,N}^T$ itself.
However, we stress that this difficulty does not affect further 
diagrammatic reorganizations which are only based on the general structure 
of (\ref{cinqtreize}). 

Now, we return to sum (\ref{cinqquatre}) over graphs  
${\cal G}_{{\cal C}}$. In each graph, we insert, into the definition
(\ref{cinqdeux}) of each statistical weight $W_{\phi}({\cal C})$, 
the expression (\ref{cinqtreize}) 
of Mayer coefficient ${\cal B}_{\phi}({\cal C}) = {\cal B}_{\phi,N}$ 
for a cluster ${\cal C}$ made with $N$ loops. The whole series  
(\ref{cinqquatre}) is then rewritten in terms of new graphs 
${\cal G}_{{\cal C}}^T$ which are still made with clusters of loops.
The weight $W_{\phi}^T({\cal C})$ of a cluster ${\cal C}$ is 
\begin{equation}
W_{\phi}^T({\cal C}) = {\prod_{{\cal L} \in {\cal C}} 
z_{\phi}({\cal L}) \over N!} 
{\cal B}_{\phi}^T({\cal C}).
\label{cinqquatorze}
\end{equation}
Like in (\ref{cinqdeux}), for root clusters ${\cal C}_a$ which contain 
the root loop ${\cal L}_a$, $N!$ is replaced by $(N-1)!$ 
in (\ref{cinqquatorze}). Two clusters are connected by at most one
bond ${\cal F}$ which reduces to either 
${\cal F}_c$, ${\cal F}_c^2/2!$ or ${\cal F}_c^3/3!$. For any one-loop 
cluster ${\cal C}={\cal L}$ which is  
doubly connected to the rest of the graph by
two bonds ${\cal F}_c$, 
weight (\ref{cinqquatorze}) is replaced by 
\begin{equation}
W_{\phi}^T({\cal L}) = z({\cal L}) (\exp (I_R({\cal L})) - 1).
\label{cinqquinze}
\end{equation}
This specific weight results from the summation of two contributions 
$W_{\phi}({\cal L})= z_{\phi}({\cal L}) = z({\cal L})\exp (I_R({\cal L}))$ and 
$z^{\ast}({\cal L}) = -z({\cal L})$ associated respectively with 
a convolution ${\cal F}_c \ast {\cal F}_c$ in a 
diagram $D_{{\cal C}}$, and from the same convolution in a graph 
${\cal G}_{{\cal C}}$.
Similarly, for a one-loop cluster ${\cal C}={\cal L}$ which is singly 
connected to the rest of the graph by one bond ${\cal F}_c^2/2!$, 
the weight $W_{\phi}^T({\cal L})$ 
also reduces to (\ref{cinqquinze}). Therefore, series (\ref{cinqquatre}) 
is exactly transformed into
\begin{equation}
\rho ({\cal L}_a) = \sum_{{\cal G}_{{\cal C}}^T} 
{1 \over S({\cal G}_{{\cal C}}^T)} \int \prod_{i=0}^n {\cal D}({\cal C}_i)
W_{\phi}^T({\cal C}_i)    
\Big[\prod {\cal F}_{\phi}\Big]_{{\cal G}_{{\cal C}}^T},
\label{cinqseize}
\end{equation}
where the sum runs over all unlabelled topologically different graphs 
${\cal G}_{{\cal C}}^T$. Like (\ref{cinqtreize}), this sum 
is restricted to graphs which become not integrable with 
respect to the relative distances between clusters when 
$\phi$ is replaced by $V$. The symmetry factor 
$S({\cal G}_{{\cal C}}^T)$ is the number of permutations of labelled clusters that leave the product of bonds   
$[\prod {\cal F}_{\phi}]_{{\cal G}_{{\cal C}}^T}$ unchanged (only clusters 
with identical numbers of loops are permuted, except the root cluster
${\cal C}_0 = {\cal C}_a$). Like in (\ref{cinqquatre}), 
the permutations of internal loops inside a given cluster are accounted 
for by the factor $1/N!$ in definition (\ref{cinqquatorze}) of
cluster weight $W_{\phi}^T({\cal C})$. This rule follows 
from the same combinatorial properties (see Appendix A) 
that have been used in the resummation process of section 3.2 or in the construction 
of graphs ${\cal G}_{{\cal C}}$. Here, such properties allow us to transform 
the sums (\ref{cinqtreize}) over labelled clusters (with labelled 
internal loops) into sums over unlabelled objects. The first graphs in (\ref{cinqseize}) are shown in Fig.10.
\begin{figure}[h]
\myfigure
{\psfrag{z*}{$z^*$}
\includegraphics{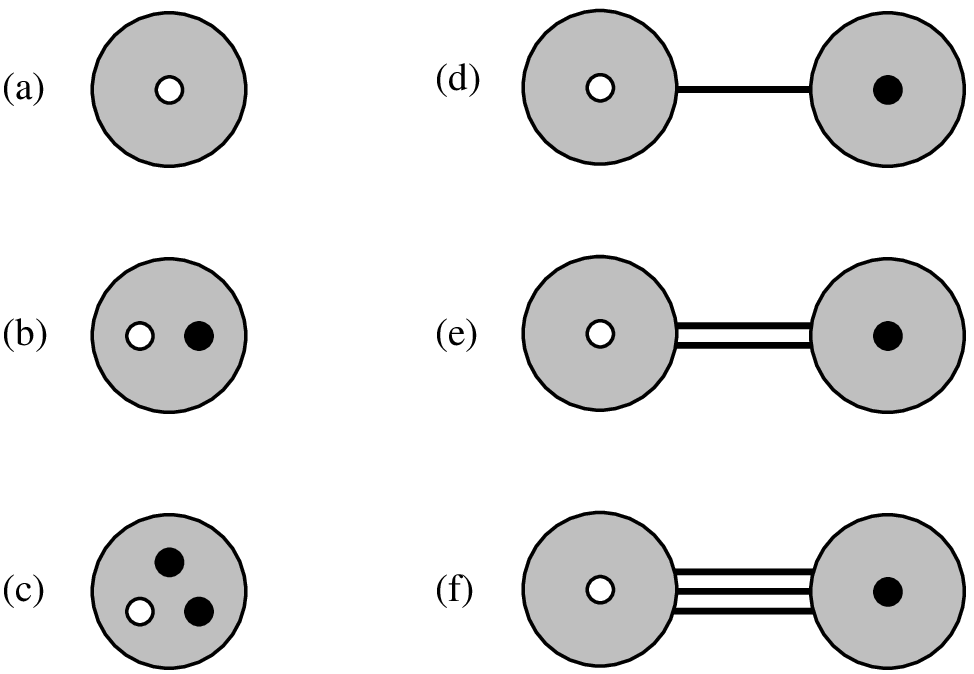}}
{	\label{fig9}
A few graphs ${\cal G}_{\CC}^T$ with one or two clusters. The bonds $\FF_\phi$ between two clusters can be either $\FF_c=-\beta\Phi$ (single line), $\frac{1}{2!}\FF_c^2$ (double line) or $\frac{1}{3!}\FF_c^3$ (triple line). The small while point represents root loop $\LL_a$. The clusters are drawn as shaded circles comprising a certain number of loops. The statistical weight of a cluster is given by the truncated weight $W_\phi^T(\CC)$ \eqref{cinqquatorze}.
}
{c}
\end{figure}

As required, each cluster weight $W_{\phi}^T({\cal C})$ remains 
integrable with respect to relative distances between its 
internal loops when $\phi$ is replaced by $V$. Notice that
products of Boltzmann factors embedded in $W_{\phi}^T({\cal C})$ are
complete: because of the structure of decomposition (\ref{cinqtreize}), 
the truncated coefficients ${\cal B}_{\phi,N}^T$, 
like the ${\cal B}_{\phi,N}$'s, do not involve incomplete 
products. Moreover, the introduction of
specific weight (\ref{cinqquinze}) can be traced back 
to the exclusion rules for prototype graphs ${\cal G}_P$ 
that avoid double counting.    

\subsection{Introduction of clusters made with particles}

If we insert (\ref{cinqseize}) into the expression (\ref{quatresepta}) of
proton density $\rho_p$, we find
\begin{equation}
\rho_p = \sum_{{\cal G}_{{\cal C}}^T} 
{1 \over S({\cal G}_{{\cal C}}^T)}
\sum_{q_a=1}^{\infty} \int {\cal D}(\bbox{\eta}_a) q_a 
\int \prod_{i=0}^n {\cal D}({\cal C}_i)
W_{\phi}^T({\cal C}_i)    
\Big[\prod {\cal F}_{\phi}\Big]_{{\cal G}_{{\cal C}}^T},
\label{cinqdixsept}
\end{equation}
where root loop ${\cal L}_a$ inside root cluster ${\cal C}_a$ in 
each graph ${\cal G}_{{\cal C}}^T$, is made with $q_a$ protons. When the 
phase space integrations over loops inside each cluster ${\cal C}_i$
are explicited, according to (\ref{quatrequatreb}), 
in ${\cal D}({\cal C}_i)$ given by (\ref{cinqcinq}), each graph 
${\cal G}_{{\cal C}}^T$ provides an infinite sum of contributions 
where clusters $C(M_p,M_e)$, made with $M_p$ protons  
and $M_e$ electrons, appear. In general, in each individual contribution, the 
weight associated with particle cluster $C(M_p,M_e)$ is not 
correctly symmetrized according to Fermi statistics. For instance, 
the graph in Fig.10c. provides, for 
$(\alpha_a = p, q_a=1;\alpha_1=e,q_1=1;\alpha_2=e,q_2=1)$, a cluster $C(1,2)$ 
with a weight that 
does not include any exchange between the two electrons. However, 
the graph in Fig.10b also provides, 
for $(\alpha_a = p, q_a=1;\alpha_1=e,q_1=2)$,
the same cluster $C(1,2)$: when both contributions are summed together, 
Fermi symmetrization is restored. In the following, we rearrange the 
whole series in terms of new graphs $G$ made with particle 
clusters that incorporate Fermi statistics.

Let us consider the class of graphs ${\cal G}_{{\cal C}}^T$ 
made with the same set of loop clusters 
$\{ {\cal C}_i \}$ and the same topological connecting structure 
of bonds ${\cal F}_{\phi}$. Two graphs in a given class differ by the number 
of internal loops in at least one cluster ${\cal C}_i$. For instance, the 
graphs in Figs. 10b. and 10c. belong to the same class defined by 
a single root cluster ${\cal C}_a$. On another hand, the graphs shown in Figs.10d., 10e. and 10f. belong to different classes because respective 
bonds ${\cal F}_{\phi}$ between clusters ${\cal C}_a$ and ${\cal C}_1$ are
different. For each cluster ${\cal C}$, 
specification of the nature and of the number of particles in all internal
loops, leads to the introduction of a particle cluster made with
$N_p$ protons and $N_e$ electrons. In each given class, 
once all graphs 
${\cal G}_{{\cal C}}^T$ have been 
replaced by infinite sums over loop particle numbers, we sum together    
the contributions associated with identical sets of particle clusters 
$\{ C_i \}$ (all respective numbers $(N_i^{(p)},N_i^{(e)})$ are identical), connected by the same bonds  ${\cal F}_{\phi}$. This defines a graph
$G$.

In a given graph $G$, each particle cluster $C(N_p,N_e)$ 
can be traced back to a loop cluster ${\cal C}$ with $N$ loops. 
In the sums $\sum_{\alpha_i} \sum_{q_i}$ obtained by expliciting 
${\cal D}({\cal C})$, there are several terms which can be associated with 
particle cluster $C(N_p,N_e)$. Each of them involves $L_p$ protonic 
loops ($\alpha_i = p$) and $L_e = N - L_p$ electronic loops ($\alpha_i = e$), while the corresponding ordered sets of particles numbers, $[q_1^{(\alpha)},...,q_{L_{\alpha}}^{(\alpha)}]$ with 
$q_1^{(\alpha)} \geq q_2^{(\alpha)} \geq ...\geq q_{L_{\alpha}}^{(\alpha)}$, 
are such that
\begin{equation}
\sum_{k=1}^{L_{\alpha}} q_k^{(\alpha)} = N_{\alpha}.
\label{cinqdixhuit}
\end{equation}
The internal state of a particle cluster $C$ then appears to be defined by 
the partitions $Q_{\alpha} = [q_1^{(\alpha)},...,q_{L_{\alpha}}^{(\alpha)}]$ 
of $N_{\alpha}$, and by the internal degrees of freedom of 
$L_{\alpha}$ loops with particles numbers determined by  
$Q_{\alpha}$, $L_{\alpha}$ being in the range $1,...,N_{\alpha}$. The corresponding cluster of loops is 
denoted by ${\cal C}(Q_p,Q_e)$. An example of all possible partitions 
$Q_{\alpha}$ and clusters of loops ${\cal C}(Q_p,Q_e)$ associated with 
the same particle cluster $C(3,2)$, are shown in Fig.11.
\begin{figure}[h]
\myfigureNoLine
{
\setlength{\extrarowheight}{1mm}
\begin{tabular}{
|>{\raggedright}m{1.5cm}
>{\raggedright}m{1.3cm}
|>{\hspace{3mm}}m{5cm}|
}
\hline
\multicolumn{3}{|c|}{\bf Particle cluster $C(3,2)$} \\
\hline
\hline
    \ $Q_p$	&	\ $Q_e$	&	\multicolumn{1}{c|}{$\CC(Q_p,Q_e)$} 	 	\\
\hline
    $[3]$		&	$[2]$	&	\includegraphics{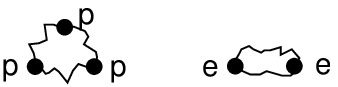}	\\
    $[3]$		&	$[1,1]$	&	\includegraphics{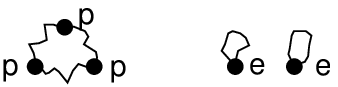}	\\
    $[2,1]$	&	$[2]$	&	\includegraphics{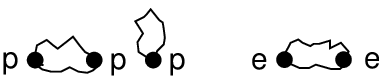}	\\
   $[2,1]$		&	$[1,1]$	&	\includegraphics{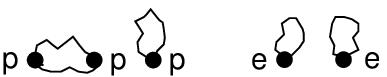}	\\
    $[1,1,1]$	&	$ [2]$	&	\includegraphics{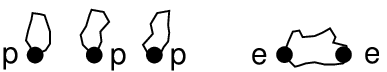}	\\
    $[1,1,1]$	&	$[1,1]$	&	\includegraphics{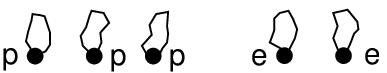}	\\
\hline
\end{tabular}
}
{	\label{fig10}
The six possible partitions $(Q_p,Q_e)$, and the associated set of loops $\CC(Q_p,Q_e)$, involved in the internal state of the particle cluster $C(3,2)$ made of 3 protons and 2 electrons.
}
{c}
\end{figure}

The statistical weight $Z_{\phi}^T(C)$ depends on internal 
state $(Q_p, Q_e; {\cal C}(Q_p,Q_e))$ of $C$. In the sum 
$\sum_{\alpha_i}$ arising from ${\cal D}({\cal C})$, there are 
$N!/(L_p!L_e!)$ terms which define 
a loop cluster ${\cal C}(Q_p,Q_e)$ made with $L_p$ protonic loops 
and $L_e$ electronic loops. In the corresponding sum $\sum_{q_i}$, 
there are $L_p!L_e! / (\prod_{q=1}^{\infty} n_p(q)! n_e(q)!)$ terms 
associated with the same partitions $Q_p$ and $Q_e$, where 
$n_{\alpha}(q)$ is the number of loops with $q$ particles of species $\alpha$. Therefore there are 
\begin{equation}
{N! \over L_p! L_e!} \times {L_p! L_e! \over 
\prod_{q=1}^{N_p} n_p(q)! \prod_{q=1}^{N_e} n_e(q)!} = 
{N! \over \prod_{q=1}^{\infty} n_p(q)! \prod_{q=1}^{\infty} n_e(q)!}
\label{cinqdixneuf} 
\end{equation}
different terms in  $\sum_{\alpha_i} \sum_{q_i}$ which contribute to  
$Z_{\phi}^T(C)$. Once integrations over positions and shapes of loops 
are performed, all these terms provide identical contributions. Thus, 
we define $Z_{\phi}^T(C)$ as the product of the number of terms 
(\ref{cinqdixneuf}) by the weight $W_{\phi}^T({\cal C}(Q_p,Q_e))$ of 
loop cluster ${\cal C}(Q_p,Q_e)$. The phase space measure 
${\cal D}(C)$ involves all possible internal states of $C(N_p,N_e)$: it
reduces to a discrete sum over all partitions $(Q_p,Q_e)$ 
satisfying the total particle number constraints (\ref{cinqdixhuit}), 
combined to spatial and functional integrations over positions and shapes 
of loops inside ${\cal C}(Q_p,Q_e)$.

The common connecting structure of graphs ${\cal G}_{{\cal C}}^T$ remains 
unchanged through the summations leading to graphs $G$. Two particle 
clusters $C_i$ and $C_j$ are connected by at most one bond ${\cal F}_{\phi}$ 
which can be either $-\beta \Phi$, $\beta^2 \Phi^2/2!$, $-\beta^3 \Phi^3/3!$; 
here $\Phi$ is the total interaction potential between two loop clusters 
${\cal C}_i(Q_i^{(p)},Q_i^{(e)})$ and 
${\cal C}_j(Q_j^{(p)},Q_j^{(e)})$ that describe 
internal states of $C_i$ and $C_j$ respectively. Again, by virtue of 
the combinatorial properties derived in Appendix A, 
$S({\cal G}_{{\cal C}}^T)$ is exactly transformed into  
$S(G)$ which is the number of permutations of clusters $C_i$ 
(except $C_a$) with identical particle numbers $(N_i^{(p)},N_i^{(e)})$ 
that leave the product of bonds ${\cal F}_{\phi}$ unchanged. Similarly to 
(\ref{cinqdixsept}), the resulting series only involve graphs $G$ 
which are no longer integrable 
over the relative distances between the particle clusters 
when $\phi$ is replaced by $V$. 

\subsection{The screened cluster expansion}

According to section 4.3, diagrammatic series (\ref{cinqdixsept}) is 
exactly transformed into the so called screened cluster expansion, i.e.
\begin{equation}
\rho_p = \sum_{G} 
{1 \over S(G)}
\int {\cal D}(C_a) Z_{\phi}^T(C_a) q_a 
\int \prod_{i=1}^n {\cal D}(C_i) Z_{\phi}^T(C_i)
\Big[\prod {\cal F}_{\phi}\Big]_{G} .
\label{cinqvingt}
\end{equation} 
The graphs $G$ are identical to the usual Mayer graphs, where the points 
are now particle clusters, except for some specific rules.
The rules which define the graphs  $G$ are listed as follows:

\bigskip

\noindent $\bullet$ Every graph $G$ involves 
the root cluster $C_0 = C_a$ and $n$ ($n \geq 0$) black clusters 
$(C_1,...,C_n)$. The root cluster $C_a$ contains the 
root proton which is fixed at the origin in $\rho_p$. Each cluster 
$C_i$ ($i=0,...,n$) contains $N_i^{(p)}$ protons and $N_i^{(e)}$ electrons.

\bigskip

\noindent $\bullet$ The internal state of a cluster $C(N_p,N_e)$ 
($C \in \{C_i, i=0,...,n \}$) 
is determined, on one hand, by partitions  
$Q_{\alpha} = [q_1^{(\alpha)},...,q_{L_{\alpha}}^{(\alpha)}]$ with  
$q_1^{(\alpha)} \geq q_2^{(\alpha)} \geq ...\geq q_{L_{\alpha}}^{(\alpha)}$ 
($\alpha=p,e$); 
on another hand, by loop cluster 
${\cal C} (Q_p,Q_e)$ made with $L_{\alpha}$ loops 
$({\cal L}_1^{(\alpha)},..., {\cal L}_{L_{\alpha}}^{(\alpha)})$ where 
${\cal L}_k^{(\alpha)}$ carries $q_k^{(\alpha)}$ particles of 
species $\alpha$ ($\alpha=p,e$). The numbers $n_{\alpha}(q)$ of loops carrying $q$ 
particles of species $\alpha$ satisfy the total particle number constraints  
\begin{equation}
\sum_{q=1}^{N_{\alpha}} q n_{\alpha}(q) = N_{\alpha}.
\label{cinqvingtdeux}
\end{equation}

\bigskip

\noindent $\bullet$ The phase space measure ${\cal D}(C)$ for a cluster 
$C(N_p,N_e)$ reads
\begin{equation}
{\cal D}(C) = \sum_{Q_p,Q_e} \int \prod_{k=1}^{L_p} \dd \textbf{X}_k^{(p)} 
\prod_{k=1}^{L_e} \dd \textbf{X}_k^{(e)}
\int \prod_{k=1}^{L_p} {\cal D}(\bbox{\eta}_k^{(p)})
\prod_{k=1}^{L_e} {\cal D}(\bbox{\eta}_k^{(e)}) 
\label{cinqvingttrois}
\end{equation}
where $\textbf{X}_k^{(\alpha)}$ and 
$\bbox{\eta}_k^{(\alpha)}$ denote 
position and shape of loop ${\cal L}_k^{(\alpha)}$ 
respectively. The discrete sum is carried out over all partitions 
$(Q_p,Q_e)$ satisfying (\ref{cinqvingtdeux}).
For root cluster $C_a$, loop ${\cal L}_1^{(p)}$ is identified 
to root loop ${\cal L}_a$ ($q_1^{(p)} = q_a$), and no integration is performed over 
position $\textbf{X}_1^{(p)} = \textbf{X}_a$ which is fixed at the 
origin.

\bigskip

\noindent $\bullet$ The statistical weight $Z_{\phi}^T(C)$ for a cluster 
$C(N_p,N_e)$ reads 
\begin{equation}
Z_{\phi}^T(C) = {\prod_{k=1}^{L_p} z_{\phi}({\cal L}_k^{(p)}) 
\prod_{k=1}^{L_e} z_{\phi}({\cal L}_k^{(e)}) \over 
\prod_{q=1}^{N_p} n_p(q)! \prod_{q=1}^{N_e} n_e(q)!} 
{\cal B}_{\phi}^T({\cal C}(Q_p,Q_e)).
\label{cinqvingtquatre}
\end{equation}
For root cluster $C_a$, $n_p(q_a)!$ is replaced 
by $(n_p(q_a)-1)!$. For a cluster $C$ different from $C_a$, 
the internal state of which 
is determined by a single protonic loop ${\cal L}_1^{(p)}$ 
or a single electronic loop ${\cal L}_1^{(e)}$, 
the expression (\ref{cinqvingtquatre}) is replaced by ($\alpha=p, e$)
\begin{equation}
Z_{\phi}^T(C) = z_{\phi}({\cal L}_1^{(\alpha)}) - z({\cal L}_1^{(\alpha)}) 
\label{cinqvingtquatrea}
\end{equation}
when $C$ is either, the intermediate cluster of a convolution 
${\cal F}_{\phi} \ast {\cal F}_{\phi}$, or connected to the rest 
of the graph by a single bond ${\cal F}_{\phi}^2/2!$. 
The truncated Mayer coefficient ${\cal B}_{\phi,N}^T$ is 
defined by a suitable truncation (see Section 4.2) 
of the usual Mayer coefficient 
${\cal B}_{\phi,N}$ for $N$ loops with pair interactions $\phi$. This 
truncation ensures that ${\cal B}_{\phi,N}^T$ remains integrable over 
relative distances between loops when $\phi$ is replaced 
by $V$. The first truncated Mayer coefficients are
\begin{equation}
{\cal B}_{\phi,1}^T = 1, \qquad
{\cal B}_{\phi,2}^T = \exp (-\beta \phi) - 1 + \beta \phi 
-{\beta^2 \phi^2 \over 2!} + {\beta^3 \phi^3 \over 3!}, \qquad
...
\label{cinqvingtcinq}
\end{equation}

\bigskip

\noindent $\bullet$ Two clusters $C_i$ and $C_j$ are connected by at most one bond 
${\cal F}_{\phi}(C_i,C_j)$ which can be either 
$-\beta \Phi$, $\beta^2 \Phi^2/2!$, $-\beta^3 \Phi^3/3!$. 
The potential $\Phi(C_i,C_j)$ 
is nothing but the total interaction potential between loop clusters 
${\cal C}_i(Q_i^{(p)},Q_i^{(e)})$ and 
${\cal C}_j(Q_j^{(p)},Q_j^{(e)})$ that describe 
internal states of $C_i$ and $C_j$ respectively, i.e.
\begin{equation}
\Phi(C_i,C_j) = \Phi({\cal C}_i,{\cal C}_j) = \sum_{\LL\in\CC_i}\sum_{\LL'\in\CC_j} \phi(\LL,\LL').
\label{cinqvingtsix}
\end{equation}
The product of all bonds inside $G$ is denoted by 
$[\prod {\cal F}_{\phi}]_{G}$. The graph $G$
is connected, i.e. for any pair $(C_i,C_j)$ there exists at least 
one connecting path made with one or more bonds ${\cal F}_{\phi}$.

\bigskip

\noindent $\bullet$ The symmetry factor $S(G)$ is the number of permutations 
of labelled black clusters $C_i$ ($i \geq 1$) that leave the product  
$[\prod {\cal F}_{\phi}]_{G}$ unchanged. Only clusters with identical 
numbers of protons and electrons are permuted. 

\bigskip

\noindent $\bullet$ The sum $\sum_{G}$ runs over all topologically different 
unlabelled graphs $G$ which are no longer integrable over 
relative distances between clusters $\{C_i, i=0,...,n \}$ when 
$\phi$ is replaced by~$V$. 

\bigskip 
\begin{figure}[h]
\myfigure
{\includegraphics{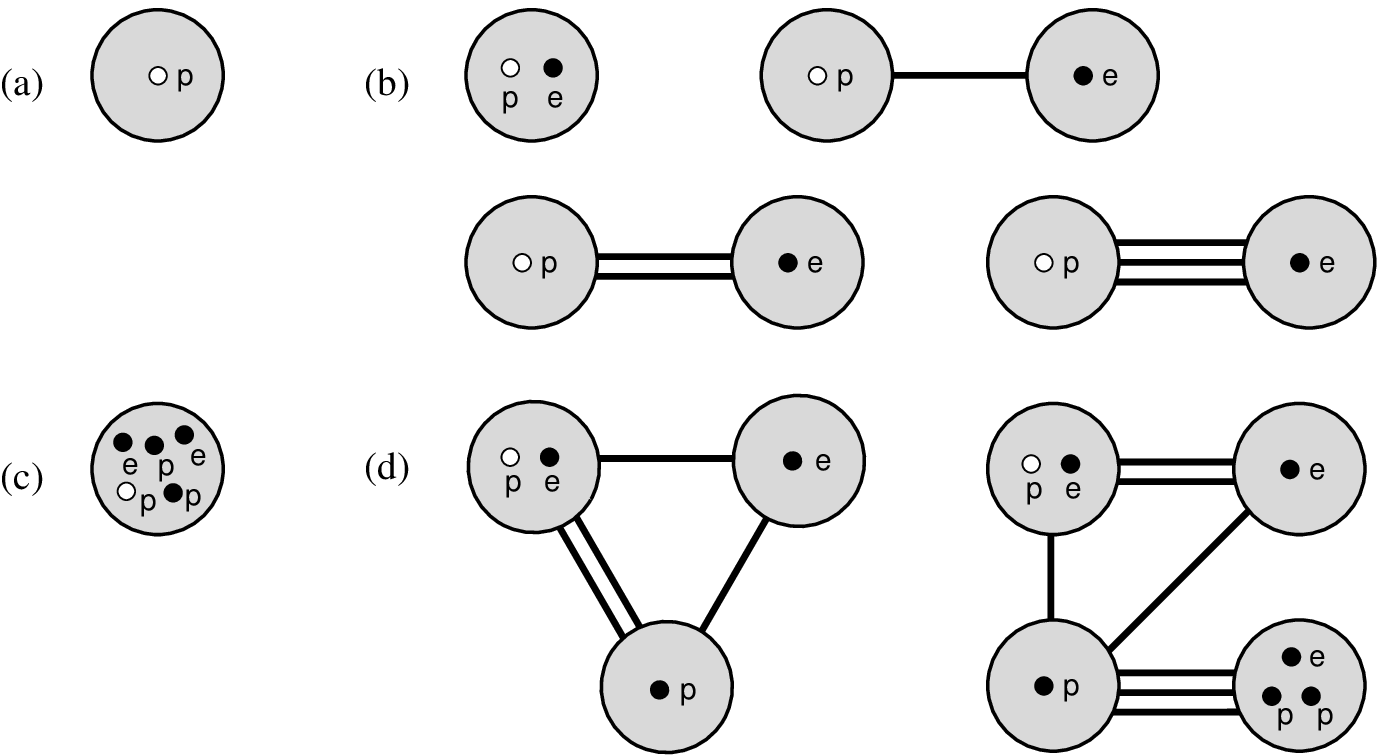}}
{	\label{fig11}
A few graphs $G$ occuring in the screened cluster expansion of $\rho_p$. (a) graph made of a single proton; (b) graphs involving a proton and an electron; (c) graph consisting in the cluster $C(3,2)$, whose possible internal states are listed in Fig.~\ref{fig10}; (d) further examples.
}
{c}
\end{figure}

Several examples of graphs $G$ are shown in Fig.12.
We stress that graphs $G$ have the same topological structure as 
familiar Mayer diagrams: the ordinary points are replaced by particle 
clusters, while usual Mayer links are now bonds ${\cal F}_{\phi}$.
The introduction of specific weight (\ref{cinqvingtquatrea}) avoids 
double counting arising from elimination of $V$ in favor of $\phi$. For 
a system with short range forces, a diagrammatic series similar to 
(\ref{cinqvingt}) can be derived without any prior chain resummations: 
$\phi$ is merely replaced by the genuine potential, and all the above
rules are conserved except that associated with specific 
weight (\ref{cinqvingtquatrea}). Of course, in this case, the sum of all graphs 
with the same number $N$ of particles can be rewritten in terms of 
the sole standard Mayer coefficient of order $N$. Here, the (finite) sum 
of all graphs $G_{N_p,N_e}$ with the same total numbers 
$N_p$ of protons and $N_e$ of electrons, 
provides a contribution, that can be rewritten as 
$(z_p^{N_p}z_e^{N_e}/(N_p-1)!N_e!) B_{N_p,N_e}^{(R)}$ by definition of
renormalized Mayer coefficient $B_{N_p,N_e}^{(R)}$. Then, 
screened cluster expansion (\ref{cinqvingt}) can be recast as 
\begin{equation}
\rho_p = \sum_{N_p=1,N_e=0}^{\infty} {z_p^{N_p}z_e^{N_e} \over (N_p-1)!N_e!} B_{N_p,N_e}^{(R)}. 
\label{sixun}
\end{equation}
The long range divergencies which plague the bare coefficient $B_{N_p,N_e}$,
(see Section 2.2) are removed: every graph $G_{N_p,N_e}$ 
is finite because $\Phi (C_i,C_j)$ 
decays as a dipolar interaction for large separations between $C_i$ and $C_j$.
Since screened potential $\phi$ depends on fugacities, 
$B_{N_p,N_e}^{(R)}$ does not depend on the sole temperature, 
contrary to usual Mayer coefficients for
systems with short range forces.

\subsection{Particle correlations and other observables}

The particle correlations, as well as other physical observables, 
can be inferred from multi-body loop densities. 
Hence, the screened cluster expansion 
of such physical quantities will directly follow from the expansions of 
these loop densities that are analogous to (\ref{cinqvingt}). 
In particular, by starting 
from (\ref{cinqseize}) and following the same lines as in Section 4.3, 
we easily obtain for (one-body) loop density $\rho(\LL_a)$ 
\begin{equation}
\rho(\LL_a) = \sum_G \frac{1}{S(G)}\int\DD(C_a) Z_\phi^T(C_a)
\Big(\sum_{\LL\in C_a} \delta(\LL,\LL_a) \Big)
\int\prod_{i=1}^n \DD(C_i) Z_\phi^T(C_i) \Big[\prod\FF_\phi\Big]_G,
\label{cinqvingtsept}
\end{equation}
where graphs $G$ are identical to those in (\ref{cinqvingt}). The sole 
difference between (\ref{cinqvingtsept}) and (\ref{cinqvingt}) lies 
in the integration upon degrees of freedom of root cluster 
$C_a$. Now $\DD(C_a)$ and $Z_\phi^T(C_a)$ are also  
given by (\ref{cinqvingttrois}) and (\ref{cinqvingtquatre}) respectively, 
while factor
$\delta(\LL,\LL_a)$ selects the contribution of 
internal states of $C_a$ where one loop $\LL$ is identified 
to $\LL_a$. It is easily checked that 
series \eqref{cinqvingt} for the protonic density 
can be recovered by inserting (\ref{cinqvingtsept}) into \eqref{quatresepta}. 
Notice that factor $1/(n_p(q_a)-1)!$ occuring in \eqref{cinqvingt} 
results from the fact that there are $n_p(q_a)$ loops in \eqref{cinqvingtsept} identified to $\LL_a$. The screened cluster expansion of 
other one-body physical quantities might be also merely derived from (\ref{cinqvingtsept}).

The truncated two-body loop density (loop correlations), 
can be expanded similarly to \eqref{cinqvingtsept}. 
Its screened cluster expansion reads
\begin{multline}
\rho_{\text{{\tiny T}}}(\LL_a,\LL_b) = \sum_G \frac{1}{S(G)}\int\DD(C_{ab}) \Big( \sum_{\LL_k,\LL_l \in \CC_{ab}, k \neq l} \delta(\LL_k,\LL_a) \delta(\LL_l,\LL_b) \Big) Z_\phi^T(C_{ab}) \\
\times \int\prod_i \DD(C_i) Z_\phi^T(C_i) \Big[\prod\FF_\phi\Big]_G + \\
+ \sum_G \frac{1}{S(G)}\int\DD(C_a)\DD(C_b) \Big( \sum_{\LL_k \in \CC_a} \delta(\LL_k,\LL_a) \sum_{\LL_l \in \CC_b} \delta(\LL_l,\LL_b) \Big)  Z_\phi^T(C_a) Z_\phi^T(C_b) \\
\times \int\prod_i \DD(C_i) Z_\phi^T(C_i) \Big[\prod\FF_\phi\Big]_G .
\label{cinqvingthuit}
\end{multline}
Now, graphs $G$ involve, either a single root cluster $C_{ab}$ 
that contains both loops $\LL_a$ and $\LL_b$, or 
two roots clusters $C_a$ and $C_b$ with $\LL_a$ in $C_a$ and $\LL_b$ in $C_b$.
The rules that define such graphs with $n$ black 
clusters $\{ C_i \}$ are identical to that of (\ref{cinqvingt}). In particular, 
the graphs are connected, and only permutations of black clusters $C_i$ 
with identical numbers $N_i^{(p)}$ and $N_i^{(e)}$ are considered in 
the definition of symmetry factors. In (\ref{cinqvingthuit}), the measure
and weights associated with root clusters are identical to 
that of black clusters, as in (\ref{cinqvingtsept}).

\section{Physical interpretation of graphs $G$ and concluding remarks}

Expansion (\ref{sixun}) is a reorganization of formal series (\ref{troiscinq}) that is suitable for studying Coulomb gas at low fugacities and low temperatures, whether in a fully or partially ionized phase. Indeed, graphs in that expansion admit a natural interpretation in terms of the recombined entities and their mutual interactions, as discussed below.

\bigskip

\noindent $\bullet$ \textsl{Graphs with a single cluster}

\bigskip

\noindent First, we consider graph $G_{N_p,N_e}$ with a single root cluster $C_a$.  When fugacities go to zero, $\phi$ reduces to $V$, and we can replace all ring factors $\exp (I_R)$ by $1$ and all 
Mayer coefficients ${\cal B}_{\phi}^T$ by ${\cal B}^T$, in $Z_{\phi}^T(C_a)$. 
Then, integration over 
all internal states of $C_a$ reduces, after applying backwards 
Feynman-Kac formula, to the trace of a suitably truncated Mayer 
operator $[\exp (-\beta H_{N_p,N_e})]_{\text{Mayer}}^T$. For $G_{1,1}$ 
shown in Fig.13a, using 
expression (\ref{cinqsix}) of ${\cal B}_{\phi,2}^T$, we find  
\begin{multline}
[\exp (-\beta H_{1, 1})]_{\text{Mayer}}^T = \exp (-\beta H_{1, 1}) - 
\exp (-\beta H_{1, 1}^{(0)}) \\
+ \int_0^{\beta} \dd \tau_1 \exp [-(\beta -\tau_1)H_{1, 1}^{(0)}] V_{1, 1} 
\exp [-\tau_1 H_{1, 1}^{(0)}] \\
-\int_0^{\beta} \dd \tau_1 \int_0^{\tau_1} \dd \tau_2 
\exp [-(\beta -\tau_1)H_{1, 1}^{(0)}] V_{1, 1} 
\exp [-(\tau_1 - \tau_2) H_{1, 1}^{(0)}]  V_{1, 1} 
\exp [- \tau_2 H_{1, 1}^{(0)}] \\
+\int_0^{\beta} \dd \tau_1 \int_0^{\tau_1} \dd \tau_2 \int_0^{\tau_2} \dd \tau_3
\exp [-(\beta -\tau_1)H_{1, 1}^{(0)}] V_{1, 1} 
\exp [-(\tau_1 - \tau_2) H_{1, 1}^{(0)}] \\
\times  V_{1, 1} \exp [- (\tau_2 - \tau_3) H_{1, 1}^{(0)}] V_{1, 1} 
\exp [- \tau_3 H_{1, 1}^{(0)}],
\label{sixdeux}
\end{multline} 
where $H_{1,1}^{(0)}=H_{1,0} + H_{0,1}$ is the kinetic part of $H_{1,1}$, while 
$V_{1,1}$ is its potential part. Similar expressions for   
$[\exp (-\beta H_{N_p,N_e})]_{\text{Mayer}}^T$ can be obtained from the definition of 
${\cal B}_{\phi,N}^T$ (see Section 4.2). In the resulting
truncated structure analogous to (\ref{sixdeux}), each Gibbs factor 
is associated with a complete Hamiltonian $H_{M_p,M_e}$ ($M_p \leq N_p$ and $M_e \leq N_e$) 
that does incorporate any pair interactions between $M_p$ protons 
and $M_e$ electrons. At low fugacities, the leading contribution 
of $G_{N_p,N_e}$ to $B_{N_p,N_e}^{(R)}$ then reads
\begin{equation}
{1 \over \Lambda} \Tr \left[\exp (-\beta H_{N_p,N_e})\right]_{\text{Mayer}}^T.
\label{sixtrois}
\end{equation}
Truncation in $[\exp (-\beta H_{N_p,N_e})]_{\text{Mayer}}^T$, inherited from that 
in the ${\cal B}^T$'s, ensures that trace (\ref{sixtrois}) 
is finite contrary to that of 
$[\exp (-\beta H_{N_p,N_e})]_{\text{Mayer}}$ defining $B_{N_p,N_e}$: when it is expressed in position and spin spaces,  
matrix elements of $[\exp (-\beta H_{N_p,N_e})]_{\text{Mayer}}^T$ are 
integrable for any spatial configuration, i.e. their decay is sufficiently 
fast with respect to any large particle separation. Moreover, that trace 
is correctly symmetrized with respect to Fermi statistics as a direct consequence of summation $\sum_{Q_p,Q_e}$ in ${\cal D}(C_a)$ 
over all possible distribution of particles into loops. 

\noindent According to the above properties, trace (\ref{sixtrois}) 
does account for recombination into complex entities. Indeed, when  
temperature goes to zero, Gibbs operators associated with $H_{M_p,M_e}$ provide 
exponentially growing terms arising from 
possible bound states with negative ground state energies $E_{M_p,M_e}^{(0)}$. 
Single term ($M_p=N_p,M_e=N_e$) can be interpreted as the ideal contribution, 
of the possible complex entity made of $N_p$ protons 
and $N_e$ electrons, while product of terms ($M_p < N_p, M_e < N_e$) describe 
its dissociation into several 
entities. Ideal contributions associated with 
familiar entities can be singled out in various graphs. For instance, $G_{1,1}$ shown in Fig.13a provides the ideal contribution
\begin{equation}
4z_pz_e \big[(m_p + m_e)/(2 \pi \beta \hbar^2)\big]^{3/2} \exp(-\beta E_H),
\label{sixquatre}
\end{equation}
of atoms $H$ in their groundstate, or $G_{2,2}$ shown in Fig.13b provides the ideal contribution
\begin{equation}
2 z_p^2 z_e^2 \big[{(2m_p + 2m_e) / (2\pi\beta\hbar^2)}\big]^{3/2} \exp (-\beta E_{H_2})
\label{sixcinq}
\end{equation}
of molecules $H_2$ in their groundstate with energy $E_{H_2}=E_{2,2}^{(0)}$. Notice that trace (\ref{sixtrois}) 
appears as the natural $(N_p+N_e)$-extension of
the so-called second virial coefficient first introduced by Ebeling 
\cite{Ebe} for $N_p=N_e=1$ (in that latter case, 
leading behaviour (\ref{sixquatre}) of (\ref{sixtrois}) can be readily 
inferred from perturbative low-temperature expansions \cite{KraKreEbeRop}).

\noindent If leading contributions to the present graph $G_{N_p,N_e}$ arise from
simple or complex entities in their ground state with bare interactions, 
next corrections involve thermal excitations and collective effects.     
For instance, in $G_{2,2}$, corrections to (\ref{sixcinq}) arise from 
excited states of molecule $H_2$, as well as 
from thermal dissociation into atoms $H$, 
ions $H_2^+$ and $H^-$, protons and electrons.  
First corrections due to screening are obtained by expanding, 
in $Z_{\phi}^T(C_a)$, all ring factors  
$\exp(I_R)$ in powers of $I_R$ and all
Mayer coefficients ${\cal B}_{\phi}^T$ in powers of
$\beta (\phi - V)$ (both $I_R$ and $\beta (\phi - V)$ 
go to zero in a sufficiently dilute regime). 
At finite fugacities, non-perturbative 
collective effects like pressure dissociation, 
are accounted for in ${\cal B}_{\phi}^T$. Since $\phi$ decays 
faster than the Coulomb potential, excited bound states with typical sizes 
larger than the screening length disappear: for instance, Rydberg states of 
atom $H$ with principal quantum number $n$, are removed 
from contribution of $G_{1,1}$ (see Fig.13a) for $n^2 a_B > \kappa^{-1}$. 
Moreover, spectral broadening of energy levels in vacuum,   
directly follows from double-time dependence of 
$\phi(\LL_i,\LL_j)$ in $s_i$ and $s_j$ (any line element of 
$\LL_i$ interacts with any line element of $\LL_j$). Since the 
equal-time condition that defines $V$ (see (\ref{quatresept})) 
no longer holds for $\phi$ \cite{BalMarAla},  
functional integration of ${\cal B}_{\phi}^T$ cannot be expressed 
in terms of an effective Hamiltonian. This is 
a particular manifestation of a general property regarding the motion of 
quantum particles coupled to a thermalized bath \cite{Fey}. 

\bigskip

\noindent $\bullet$ \textsl{Graphs with two clusters and more}

\bigskip  

\noindent According to the previous analysis, a particle cluster 
$C(N_p,N_e)$ can be associated with a possible complex entity 
made with $N_p$ protons and $N_e$ electrons. Therefore, a graph $G$ 
with several clusters describes interactions between such entities. 
We consider successively various contributions arising from 
bonds ${\cal F}_{\phi}(C_i,C_j)$ between either neutral or charged 
particle clusters $C_i$ and $C_j$. 

\noindent Let us consider two neutral clusters $C_i$ and $C_j$ (numbers of protons 
and electrons in each given cluster are identical).
Even if screening is incorporated into $\Phi(C_i,C_j)$, at sufficiently low 
fugacities $\Phi(C_i,C_j)$ can be replaced by the bare interaction 
between neutral clusters. Indeed, like $\phi$ itself, that interaction 
displays a dipolar structure at large distances, and its 
contributions to $G$ remain integrable. For instance, in graph $G_{2,2}$ 
shown in Fig.13c, collective effects in $Z_{\phi}^T(C_a)$, $Z_{\phi}^T(C_1)$ 
and $\Phi(C_a,C_1)$ can be neglected in a sufficiently dilute regime:
the corresponding leading contribution 
accounts for familiar van der Waals interactions between two atoms $H$. 

\noindent Now we turn to a neutral cluster $C_i$ connected to a charged cluster 
$C_j$ by a bond $\beta^2 \Phi^2/2!$ or $-\beta^3 \Phi^3/3!$.
At sufficiently low fugacities, $\Phi(C_i,C_j)$ can be replaced, 
by the bare dipole-charge interaction between $C_i$ and $C_j$, 
the square or cube of which is indeed integrable at large distances. 
For instance in graph $G_{2,3}$ shown in Fig.13d, leading contribution 
describes polarization effects of an atom $H$ submitted to the 
electric field created by an ion $H^-$. In a bond $-\beta \Phi$, screening 
is \textsl{a priori} crucial for ensuring integrability (however, for some 
graphs sufficiently connected, leading contributions can still be 
estimated by keeping only bare interactions). 

\noindent Eventually, we consider pairs $(C_i,C_j)$ of charged clusters with 
a connecting bond ${\cal F}_{\phi}(C_i,C_j)$. Effective 
interaction $\Phi(C_i,C_j)$ in each bond takes into account 
screening due to free protons and free electrons at a mean-field level. 
Suitable combinations of charged pairs and bonds 
in $G$, provide contributions that  
account for various mechanisms. For instance, graph $G_{2,1}$ shown in Fig.13e 
gives screened contributions beyond mean-field Debye-like terms. 
Graph $G_{4,1}$ shown in 
Fig.13f describes modification of screened interactions between free protons 
due to the presence of ions $H_2^+$. 
Summation of all similar graphs with 
convolution chains $(-\beta \Phi) \ast (-\beta \Phi) \ast...\ast (-\beta \Phi)$ 
and intermediate charged complex entities, allows us to introduce effective 
interactions screened by all charges present in the medium. 
Polarization of charged complex entities by other charges also appear.     
Graph $G_{3,1}$ shown in 
Fig.13g incorporates polarization effects of an ion 
$H_2^+$ by a free proton which is itself 
screened over distances of order $\kappa^{-1}$. 

\begin{figure}[h]
\myfigure
{\includegraphics{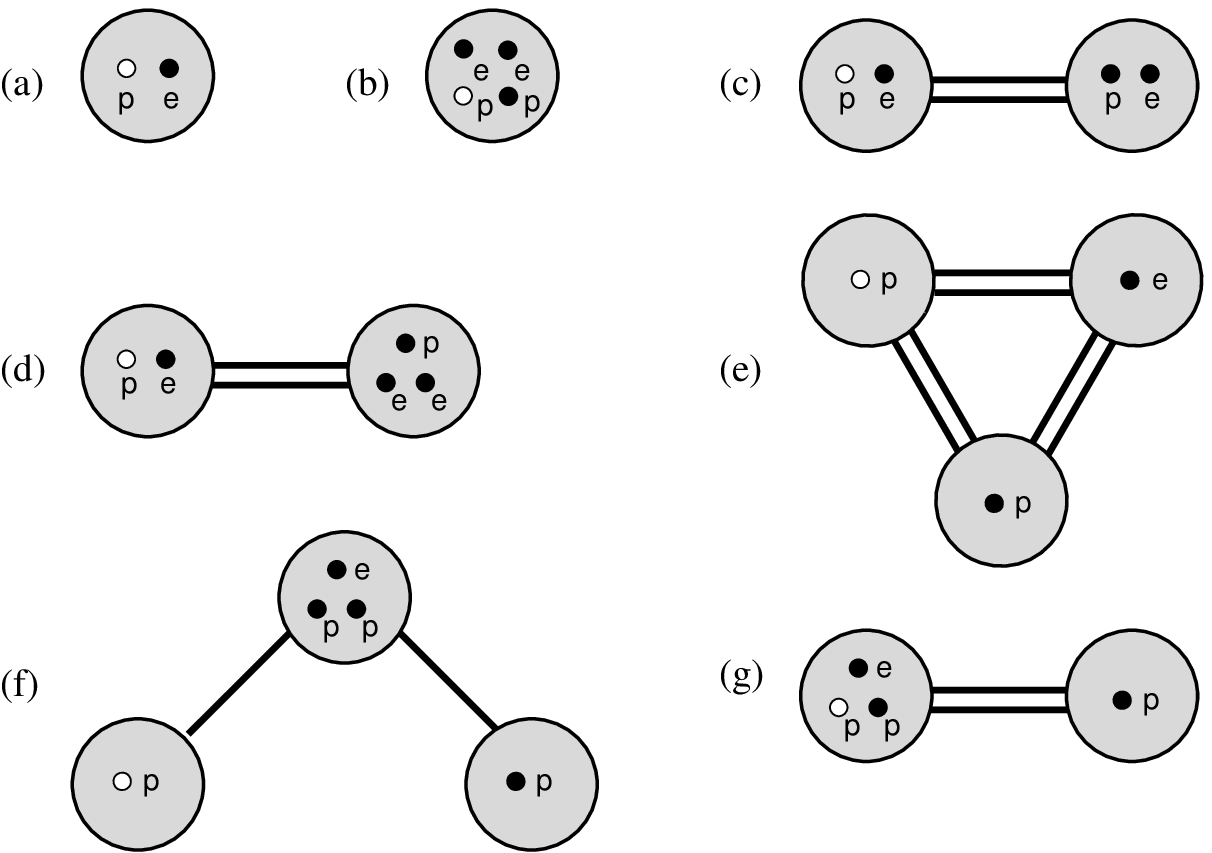}}
{	\label{fig12}
Examples of graphs $G$ describing various physical effects. Formation of bound states: (a) hydrogen atom (b) hydrogen molecule $H_2$; (c) van der Waals interactions between two H atoms; (d) polarization of atom $H$ by the field of a $H^-$ ion; (e) contribution beyond mean-field to the screened interactions between two free (ionized) protons and a free electron; (f) modification of the screened interaction between two free protons due to a $H_2^+$ ion; (g) polarization of an ion $H_2^+$ by the field of a free proton screened over distances of order $\kappa^{-1}$.
}
{c}
\end{figure}

\bigskip

Thus, screened cluster expansion (\ref{cinqvingt}) provides a complete 
framework that sustains the so-called chemical picture 
at sufficiently low temperatures and low densities. 
Any complex entity made with $N_p$ protons and $N_e$ electrons can 
be associated with a particle cluster $C(N_p,N_e)$, while both ideal and 
interaction contributions are incorporated into statistical weights  $Z_{\phi}^T(C)$ and bonds ${\cal F}_{\phi}$. 
Every physical mechanism (see examples above) is described 
by a well behaved graph $G$, and consequently it can be estimated 
in a perfectly controlled and unambiguous way. We stress that screened 
cluster expansions (\ref{cinqvingt}), (\ref{cinqvingtsept}), 
(\ref{cinqvingthuit}) can be extended to any quantum 
nucleo-electronic plasma. Then particle clusters are made with any kind of 
nuclei and electrons, while statistical weights and bonds account for 
specific masses, charges and spins of all nuclei species. All 
topological and combinatorial rules 
that define graphs $G$, remain unchanged. Usefulness of 
screened cluster formalism for studying partially ionized gases 
is illustrated by various applications as described below.

It has been rigorously shown \cite{LieColYau,MacMar} that the equation of state 
of the Hydrogen plasma
reduces to that of an ideal mixture of protons, electrons, 
and atoms $H$ in their ground state, in some suitable zero-temperature 
scaling limit defining the so-called Saha regime.
The first rigorous result on the atomic limit has been established by Fefferman \cite{Fef} (for a review, see \cite{BryMar}).   Screened cluster expansions allow us to recover immediately 
ideal terms associated with free protons, free electrons and atoms $H$.
All other contributions are shown to decay exponentially faster 
in the Saha regime. For instance, the ideal contribution (\ref{sixcinq}) 
of molecules $H_2$ is smaller than (\ref{sixquatre}) by factor 
$\exp [\beta (2\mu + E_H - E_{H_2})] \sim
\exp [\beta (3E_H - E_{H_2})]$ (for $\mu$ close to $E_H$), which indeed exponentially 
vanishes at zero temperature since $3E_H < E_{H_2}$: 
though molecule $H_2$ is more stable than two atoms $H$ ($E_{H_2} < 2E_H$), 
entropy dissociation overcomes energy minimization in the present 
very dilute regime. Moreover, 
detailed analysis of graphs $G$ provide exact estimations of various 
corrections to ideal leading terms at finite (low) temperature and density. 
Such corrections arise from excited bound states of atoms $H$, 
thermal and entropy dissociation, recombination into other complex entities 
($H_2$, $H_2^+$, $H^-$,...), screening, interactions between entities, 
diffraction and exchange (all entities behave as almost classical particles). 
Those results will be published elsewhere \cite{AlaBalCorMar}.

For the Hydrogen plasma in Saha regime, particle correlations have been 
determined \cite{AlaCorMar} by using screened cluster expansion (\ref{cinqvingthuit}). That analysis leads to the introduction 
of effective interactions between any entity, which all decay as $1/r^6$ at large relative distances $r$. In particular, 
familiar van der Waals forces between atoms $H$ do appear. At zero 
temperature, standard perturbation formula 
for two atoms in the vacuum is exactly recovered. Corrections to that 
formula due to excited states and screening, are also calculated at 
finite temperature and density.   
 
The dielectric response of the Hydrogen plasma in its atomic phase has 
been investigated with screened cluster formalism \cite{Bal,BalMar}. 
That phase is defined within previous Fefferman limit 
\cite{Fef} where $\mu$ is chosen slightly larger than $E_H$: 
it reduces to an ideal mixture of atoms $H$ 
in their groundstate (such phase may also be viewed as a particular 
Saha regime where the fraction of ionized protons and electrons 
is arbitrarily small). The behaviour of wavenumber-dependent dielectric 
constant is analyzed by using expansion (\ref{cinqvingthuit}) 
in linear response to an infinitesimal external charge. A cross-over 
phenomenon emerges, from a dielectric behaviour at wavenumbers much larger than 
$\kappa$, to a conductive behaviour at scales smaller than $\kappa$. 
Conductive behaviour is due to a finite tiny amount of free charges 
that are always present at finite temperature and density (here, 
screening length $\kappa^{-1}$ is much larger than the mean distance 
between atoms $H$). The large plateau observed  
at scales larger than $\kappa$, leads to a natural definition of the 
(finite) dielectric constant for the atomic phase, independently 
of the presence of residual free charges. At zero temperature, 
the usual expression in terms of the polarizability of a single atom 
$H$ is then exactly recovered.   

Previous examples confirm that screened cluster formalism 
is well suited for studying dilute regimes at low temperature. 
This concerns both conceptual problems for Coulomb matter, like 
the existence of van der Waals forces at finite 
temperature and finite density or the definition of a dielectric 
regime, and also derivation of exact asymptotic expressions for 
equilibrium quantities.
All those various analysis can be applied to any partially ionized 
(even neutral) gas. In addition, screened cluster expansions 
provide a promising route for deriving either 
approximation schemes or effective models. Both schemes and models 
might reasonably describe regimes with higher temperature and density, 
where analytic estimations are no longer possible.  
 
\appendix

\section{Combinatorial properties of Mayer graphs}

We derive various combinatorial properties for Mayer graphs which are used in 
the transformations described in Section 4. These properties are 
related to the following central observation. Let $G$ be an 
unlabelled Mayer graph built with $N$ internal points $x$, weights $z$ and bonds $f$ (for our purpose it is unnecessary to take explicitly into account the root points). Its contribution reads
\begin{equation}
{1 \over S(G)} \int \big[\prod \dd x \, z(x)\big]_G \big[\prod f\big]_G.
\label{A1}
\end{equation}
where $S(G)$ is its symmetry factor, i.e. the number of permutations of the 
$N$ internal points that leave unchanged the product 
of bonds $[\prod f_{ij}]_{G_{lab}}$ 
associated with a labelled diagram $G_{lab}$ with the same 
topological structure as that of $G$. The number of different labelled 
diagrams $G_{lab}$ is equal to $N!/S(G)$. Once all integrations over 
internal points $(x_1,...,x_N)$ have been performed, all contributions from different labelled diagrams are identical. Thus, (\ref{A1}) can be rewritten 
as 
\begin{equation} 
{1 \over N!} \sum_{G_{lab}} \int \prod_{i=1}^N \dd x_i \, z_i 
\big[\prod f_{ij}\big]_{G_{lab}}
\label{A2}
\end{equation}
where the sum runs over all different labelled diagrams $G_{lab}$ with the 
given topological structure determined by $G$. The transformation of 
(\ref{A1}) into (\ref{A2}) allows us to derive several useful identities as described below.

\bigskip

\noindent $\bullet$ \textsl{Decomposition of bonds}

\bigskip

\noindent If bond $f$ is decomposed into a sum, $f=g+h$, 
contribution (\ref{A1}) 
can be rewritten in terms of that of graphs $G^{\ast}$ built with 
two kinds of bonds, $g$ and $h$. Replacing each bond $f_{ij}$ 
by $g_{ij} + h_{ij}$ into (\ref{A2}), we find a sum over labelled 
diagrams $G_{lab}^{\ast}$ where the product of bonds reads 
$[\prod g_{kl}]_{G_{lab}^{\ast}}$ times $[\prod h_{mn}]_{G_{lab}^{\ast}}$. Obviously 
labelled diagrams $G_{lab}^{\ast}$ do exhibit the topological properties of 
Mayer graphs: they are connected, and two points are connected by at most 
one bond which is either $g$ or $h$. All $G_{lab}^{\ast}$'s have different products $[\prod g_{kl}]_{G_{lab}^{\ast}} [\prod h_{mn}]_{G_{lab}^{\ast}}$. 
They can be distributed into different classes where each class is defined 
by an unlabelled graph $G_{\alpha}^{\ast}$ ($\alpha=a,b,...$).
In each class, two different labelled diagrams
$G_{lab}^{\ast}$ can be inferred from each other by a permutation of 
labels of internal points, so there are $N!/S(G_{\alpha}^{\ast})$ 
such different labelled diagrams $G_{lab}^{\ast}$ which all provide the same contribution after integration over the $x$'s. Thus, we obtain
\begin{equation}
{1 \over S(G)} \int \big[\prod \dd x z(x)\big]_G \big[\prod f\big]_G = 
\sum_{G_{\alpha}^{\ast}}{1 \over S(G_{\alpha}^{\ast})} \int \big[\prod \dd x z(x)\big]_{G_{\alpha}^{\ast}} \big[\prod g\big]_{G_{\alpha}^{\ast}} 
\big[\prod h\big]_{G_{\alpha}^{\ast}}
\label{A3}
\end{equation}
where the sum runs over all different unlabelled Mayer graphs $G_{\alpha}^{\ast}$
built with two kinds of bonds $g$ and $h$. Moreover $S(G_{\alpha}^{\ast})$ is the usual symmetry factor defined as the number of permutations 
of internal points that leave the product 
$[\prod g_{kl}]_{G_{lab}^{\ast}} [\prod h_{mn}]_{G_{lab}^{\ast}}$ 
unchanged for any labelled 
diagram $G_{lab}^{\ast}$ associated with $G_{\alpha}^{\ast}$. 
The topological structures 
of graphs $G_{\alpha}^{\ast}$ are inherited from that of 
$G$: they only differ by 
number and position of bonds $g$ and $h$. If we sum (\ref{A3}) over all 
possible Mayer graphs $G$, we then find the remarkable identity
\begin{equation}
\sum_{G}{1 \over S(G)} \int \big[\prod \dd x z(x)\big]_G \big[\prod f\big]_G = 
\sum_{G^{\ast}}{1 \over S(G^{\ast})} \int \big[\prod \dd x z(x)\big]_{G^{\ast}} 
\big[\prod g\big]_{G^{\ast}} \big[\prod h\big]_{G^{\ast}}
\label{A4}
\end{equation}
where both graphs $G$ and $G^{\ast}$ are built within the same Mayer prescriptions.

\bigskip

\noindent $\bullet$ \textsl{Decomposition of weights}

\bigskip

\noindent Similarly to (\ref{A4}), we can transform (\ref{A1}) 
when weight $z$ is decomposed as 
$z=u+v$. Again in (\ref{A2}) we replace each $z_i$ by $u_i+v_i$. Following the same lines as above, we obtain the identity  
\begin{equation}
\sum_{G}{1 \over S(G)} \int\! \big[\!\prod \dd x z(x)\big]_G \big[\!\prod f\big]_G = 
\sum_{G^{\ast}}{1 \over S(G^{\ast})} \int\! \big[\!\prod \dd x\big] 
\big[\!\prod u(x)\big]_{G^{\ast}} \big[\!\prod v(x)\big]_{G^{\ast}} 
\big[\!\prod f\big]_{G^{\ast}} 
\label{A5}
\end{equation}
where now Mayer graphs $G^{\ast}$ are defined with two kinds of weights 
$u$ and $v$ (and one kind of bond $f$). The symmetry factor $S(G^{\ast})$
is defined as the number of permutations of
internal points that leave products of weights 
$[\prod u_i]_{G_{lab}^{\ast}}$, $[\prod v_j]_{G_{lab}^{\ast}}$ and 
bonds $[\prod f_{kl}]_{G_{lab}^{\ast}}$ separately unchanged. Notice that 
an identity analogous to (\ref{A5}) holds when the internal state of 
$x$ is made of two components $s$ and $t$, i.e. $\dd x= \dd s + \dd t$: 
$u(x)\dd x$ and $v(x)\dd x$ are merely replaced by $z(s)\dd s$ and
$z(t)\dd t$ respectively in the r.h.s. of (\ref{A5}).

\bigskip

\noindent $\bullet$ \textsl{Resummations of internal structures}

\bigskip

\noindent Now, let us consider labelled diagrams $G_{lab}$ 
associated with the following Mayer graph $G$. Weights can be either 
$u$ or $v$, while bonds can be either $g$, $h$ or $h^2/2!$. Two 
connected points with weights $u$ are linked by 
a bond $g$. A point with weight $v$ is either singly 
connected to a point with weight $u$ \textsl{via} a bond $h^2/2!$, 
or is the intermediate point of a convolution $h \ast h$ (the 
extremal points of convolutions $h \ast h \ast...\ast h$ carry weights 
$u$). The formula analogous to that relating (\ref{A1}) to (\ref{A2}) 
reads here
\begin{multline}
{1 \over S(G)} \int \big[\prod \dd x\big]_G \big[\prod u(x)\big]_G 
\big[\prod v(x)\big]_G \big[\prod g\big]_G \big[\prod h\big]_G \big[\prod h^2/2\big]_G  = \\
{1 \over N!} \sum_{G_{lab}} \int \prod_{i=1}^N \dd x_i 
\big[\prod u_j\big]_{G_{lab}} \big[\prod v_k\big]_{G_{lab}} 
\big[\prod g_{lm}\big]_{G_{lab}} \big[\prod h_{np}\big]_{G_{lab}} 
\big[\prod h_{qr}^2/2!\big]_{G_{lab}}. 
\label{A6}
\end{multline}
Then, we define a partition of points $x$ with weights $u$  
in terms of $M$ clusters $X$, such that the subdiagram associated with $X$ is 
connected \textsl{via} bonds $g$ only. The set of clusters $X$ is determined by 
the topological structure of the unlabelled graph $G$. Each cluster 
$X_i$ contains $N_i$ points $x$ with weights $u$, so there remains 
$L=N-\sum_{i=1}^M N_i$ points $x$ with weights $v$. There are 
$N!/(N_1!N_2!...N_M!L!M!)$ different partitions of $N$ labelled points 
into $M$ labelled clusters ($X_1,X_2,...,X_M$) and $L$ labelled 
points with weights $v$. Such labelled clusters and points (weights $v$) together with bonds $g$ and $h$, 
define a labelled diagram ${\cal G}_{lab}$. The corresponding 
$N!/(N_1!N_2!...N_M!L!M!)$ labelled diagrams $G_{lab}$
provide identical contributions in the r.h.s of (\ref{A6}). Their sum 
reduces to   
\begin{equation}
{1 \over L!M!} \int \prod_{i=1}^M \Big[{\prod_{j=1}^{N_i}\dd x_j^{(i)} u(x_j^{(i)}) 
\over N_i!} \prod g_{lm}\Big]_{X_i} \prod_{k=1}^L \dd x_k v(x_k) 
\big[\prod h_{np}\big]_{{\cal G}_{lab}} 
\big[\prod h_{qr}^2/2!\big]_{{\cal G}_{lab}}. 
\label{A7}
\end{equation}
In (\ref{A7}),   
$[\prod h_{np}]_{{\cal G}_{lab}}$ is a product of convolution chains 
$h \ast h \ast...\ast h$, with intermediate points $x_k$ (weight $v(x_k)$) 
and two ending points belonging either to the same cluster $X_i$ or 
to two different clusters $X_i$ and $X_j$. Moreover, in 
$[\prod h_{qr}^2/2!]_{{\cal G}_{lab}}$, each bond $h^2/2!$ connects 
a point $x_k$ (weight $v(x_k)$) to a point belonging to a cluster $X_i$.   
Now, we can proceed to a further summation of contributions (\ref{A7}) 
which only differ by the points 
(inside clusters $X_i$) at the extremities of convolution chains 
$h \ast h \ast...\ast h$ or bonds $h^2/2!$. This obviously leads to the 
introduction of 
\begin{equation}
H(X_i,x_k) = \sum_{j=1}^{N_i} h(x_j^{(i)},x_k), 
\label{A8}
\end{equation}
so (\ref{A6}) can be rewritten as
\begin{multline}
{1 \over S(G)} \int \big[\prod \dd x\big]_G \big[\prod u(x)\big]_G 
\big[\prod v(x)\big]_G \big[\prod g\big]_G \big[\prod h\big]_G \big[\prod h^2/2\big]_G  = \\
{1 \over L!M!} \sum_{{\cal G}_{lab}} 
\int \prod_{i=1}^M \Big[{\prod_{j=1}^{N_i}\dd x_j^{(i)} u(x_j^{(i)}) 
\over N_i!} \prod g_{lm}\Big]_{X_i} \prod_{k=1}^L \dd x_k v(x_k) 
\big[\prod H_{np}\big]_{{\cal G}_{lab}} 
\big[\prod H_{qr}^2/2!\big]_{{\cal G}_{lab}}. 
\label{A9}
\end{multline}
In (\ref{A9}), $[\prod H_{np}]_{{\cal G}_{lab}}$ is the product of convolution chains $H \ast H \ast...\ast H$ (intermediate points $x_k$ and $H(x_k, x_l) 
= h(x_k,x_l)$) that either are attached to one cluster $X_i$ or connect two clusters $X_i$ and $X_j$. Moreover, in 
$[\prod H_{qr}^2/2!]_{{\cal G}_{lab}}$, each bond $H^2/2!$ connects 
a point $x_k$ to a cluster $X_i$. The sum runs over all  
labelled diagrams ${\cal G}_{lab}$ with different products of weights and 
bonds, and the same topological structure defined by $G$. Identity 
(\ref{A9}) can be easily summed over all graphs $G$ that determine the 
following set of
labelled diagrams ${\cal G}_{lab}$: clusters $X_i$, 
points $x_k$, bonds $H_{np}$ and $H_{qr}^2/2!$, are identical, but 
internal structures of clusters in terms of bonds $g$ are different. 
Since combinatorial prefactors for each labelled diagram  
${\cal G}_{lab}$ are all identical, the sum over labelled diagrams of 
products (over $i=1,...,M$) of internal weights $[\prod g_{lm}]_{X_i}$ 
reduces to the product (over $i=1,...,M$) of sums of those weights over 
all possible internal structures.
This provides for each cluster $X_i$ the weight $Z_i$  
\begin{equation}
Z_i={\prod_{j=1}^{N_i} u(x_j^{(i)}) 
\over N_i!} \sum \big[\prod g_{lm}\big]_{X_i}        
\label{A10}
\end{equation}
where the sum runs over all different products of bonds $g_{lm}$ 
connecting labelled points inside $X_i$, i.e. over all corresponding 
Mayer labelled diagrams built with $N_i$ points $x_j^{(i)}$ and 
bonds $g_{lm}$. If we set 
\begin{equation}
\dd X_i = \prod_{j=1}^{N_i} \dd x_j^{(i)},  
\label{A11}
\end{equation}
we eventually find 
\begin{align}
\notag
\sum_G &{1 \over S(G)} \int \big[\prod \dd x\big]_G \big[\prod u(x)\big]_G 
\big[\prod v(x)\big]_G \big[\prod g\big]_G \big[\prod h\big]_G \big[\prod h^2/2\big]_G \\
\notag
&=\sum_{{\cal G}_{lab}} {1 \over L!M!}
\int \prod_{i=1}^M \dd X_i Z(X_i) \prod_{k=1}^L \dd x_k v(x_k) 
\big[\prod H_{np}\big]_{{\cal G}_{lab}} 
\big[\prod H_{qr}^2/2!\big]_{{\cal G}_{lab}} \\
&=\sum_{{\cal G}} {1 \over S({\cal G})}
\int \big[\prod \dd X Z(X)\big]_{{\cal G}}
\big[\prod \dd x v(x)\big]_{{\cal G}}
\big[\prod H\big]_{{\cal G}} 
\big[\prod H^2/2!\big]_{{\cal G}}
\label{A12}
\end{align}
where the last line follows by applying backwards the relation 
between unlabelled graphs and labelled diagrams. The permutations accounted 
for in $S({\cal G})$ act separately on sets of labelled clusters 
$(X_1,...X_M)$ and labelled points $(x_1,...,x_L)$, while they keep invariant 
each specific product of weights or bonds (this implies that permuted 
clusters necessarily contain the same number of particles).

\bigskip

All present identities result from simple combinatorics rules on one hand, 
and from the topological invariance of Mayer graphs under the considered transformations on another hand. They apply for any kind of points $x$: 
usual points (without internal degrees of freedom), loops (extended objects), 
clusters (sets of entities). Results derived in Section 4 (and those recalled in 
Section 3.2) follow by successive applications or combinations of these 
identities.\\

\paragraph{Acknowledgment}

In April 2002, one of us (AA) took part to the conference "Liquid State
Theory: from White Dwarfs to Colloids" in honour of Jean-Pierre Hansen, on 
the occasion of his 60$^{\text{th}}$ birthday. AA is deeply indebted to Jean-Pierre
for his constant support since he met him as a student, attending his
enthusiastic and pedagogical lectures. AA would like to dedicate that 
paper to Jean-Pierre, as a close friend. All authors acknowledge his 
inspired contributions to statistical mechanics of fluids, as well as his 
remarkable management of the Laboratory of Physics of the ENS lyon.

\end{document}